\documentclass[11pt,lot,lof]{duethesis}

\usepackage{amsthm}
\usepackage{amsfonts}
\usepackage{mathrsfs}
\usepackage{epsf,rotating,float,amssymb,amsmath}
\usepackage{epsfig,url,times,cite,subfigure,soul}
\usepackage{booktabs}
\usepackage{multirow}
\usepackage{graphics}
\usepackage{epstopdf}
\usepackage{gensymb}
\usepackage{amsfonts,algpseudocode}

\newtheorem{theorem}{Theorem}[section]

\newtheorem{algorithm}{Algorithm}

\newtheorem{lemma}[theorem]{Lemma}

\numberwithin{equation}{section}
\newcommand{\define}{\newcommand}
\define{\be}{\begin{eqnarray}}
\define{\ee}{\end{eqnarray}}
\define{\ben}{\begin{eqnarray*}}
\define{\een}{\end{eqnarray*}}

\title{Multi-agent Reinforcement Learning Embedded Game for the Optimization of Building Energy Control and Power System Planning}
\submitted{October, 2018}
\author{Jun Hao}
\advisor{Jun Jason Zhang}

\abstract
{
Most of the current game-theoretic demand-side management methods focus primarily on the scheduling of home appliances, and the related numerical experiments are analyzed under various scenarios to achieve the corresponding Nash-equilibrium (NE) and optimal results. However, not much work is conducted for academic or commercial buildings. The methods for optimizing academic-buildings are distinct from the optimal methods for home appliances. In my study, we address a novel methodology to control the operation of heating, ventilation, and air conditioning system (HVAC). 

We assume that each building in our campus is equipped with smart meter and communication system which is envisioned in the future smart grid. For academic and commercial buildings, HVAC systems consume considerable electrical energy and impact the personnels in the buildings which is interpreted as monetary value in this article. Therefore, we define social cost as the combination of energy expense and cost of human working productivity reduction. We implement game theory and formulate a controlling and scheduling game for HVAC system, where the players are the building managers and their strategies are the indoor temperature settings for the corresponding building. We use the University of Denver campus power system as the demonstration smart grid and it is assumed that the utility company can adopt the real-time pricing mechanism, which is demonstrated in this paper, to reflect the energy usage and power system condition in real time. For general scenarios, the global optimal results in terms of minimizing social costs can be reached at the Nash equilibrium of the formulated objective function. The proposed distributed HVAC controlling system requires each manager set the indoor temperature to the best response strategy to optimize their overall management. The building managers will be willing to participate in the proposed game to save energy cost while maintaining the indoor in comfortable zone.

With the development of Artificial Intelligence and computer technologies, reinforcement learning (RL) can be implemented in multiple realistic scenarios and help people to solve thousands of real-world problems. Reinforcement Learning, which is considered as the art of future AI, builds the bridge between agents and environments through Markov Decision Chain or Neural Network and has seldom been used in power system. The art of RL is that once the simulator for a specific environment is built, the algorithm can keep learning from the environment. Therefore, RL is capable of dealing with constantly changing simulator inputs such as power demand, the condition of power system and outdoor temperature, etc. Compared with the existing distribution power system planning mechanisms and the related game theoretical methodologies, our proposed algorithm can plan and optimize the hourly energy usage, and have the ability to corporate with even shorter time window if needed. 

Simulation results prove that the proposed methodology can set the indoor temperature with respect to real-time pricing and the number of inside occupants, maintain indoor comfort, reduce individual building energy cost and the overall campus electricity charges. Compared with the traditional game theoretical methodology, the RL based gaming methodology can achieve the optiaml resutls much more quicker.
}

\acknowledgements
{
First of all, I would like to extend my sincere gratitude to my supervisor, Dr. Jun Jason Zhang, for he instructive advice and useful suggestions on my thesis. I am deeply grateful of he help in the completion of this thesis. I learned a lot knowledge from his excellent performance in teaching and research capabilities as well as high motivation and responsibility to his students. 

I also gatefully thank Dr. Gao, Dr. Yingchen Zhang, Dr. Khodaei, whose profound knowledge of power system helped me a lot in my researches. 

I am also deeply indebted to all the other tutors and teachers for their direct and indirect help to me. 

Special thanks should go to my friends who have put considerable time and effort into their comments on the draft. 

Finally, I am indebted to my parents for their continuous support and encouragement.
}

\dedication{Dedicated to my parents.}

\begin{document}
\chapter{Chapter 1}\label{Chap1}
\section{Introduction}\label{sec:introduction01}
As society progresses and technology develops, the power system becomes more and more complicated, and humans’ requirements and expectations upon the system have been leveled up step by step. The goal of power system improvements start from successful delivery of energy, to safe energy delivery, to expansion for larger coverage and capacity, to advancements in stability and resilience, to boost of energy efficiency, and to enhancement of social welfare. Furthermore, the installation and development of renewable energy resources in most of universities or societies is speeding up, because the prices of solar panels and wind turbines are decreasing for their potential costumers so most of the buyers can afford the renewable energies nowadays. Meanwhile, The integration of renewable energy need the power grid adds more modularity and improve adaptability which may lower the system's robustness and uncertainty in terms of the balance between demand and generation. On the other hand, the integration of renewable energy will also increase the fluctuation of real-time pricing (RTP) in future smart grid, which makes the prediction and planning of distribution power system more complex. For energy-end users who consumes bulk power like universities, price is among one of the crucial factors when it comes to cost saving and load reduction \cite{SGresidential}. 

With the introduce of smart meters and nodal prices, the utility price for the buses within a same distribution power grid varies a lot. Therefore, building managers have their own control objectives according to the corresponding distribution locational marginal prices \cite{hao2016locational}. However, the nodal price in a distribution power grid is not only dependent on one bus, it is almost under the influence of every bus in the same distribution power grid. Therefore, each building manager has to take the influences from other buses into consideration before they make their decision. Therefore, within the distribution power grid, the building managers can play a game and find the optimal strategy for them to control the HVAC system. However, the outcome, generated from the game, is solely dependent on the utility cost. For large commercial buildings and academic buildings, this kind of control and planning strategy needs to be improved to ensure that the control strategy would not affect the working efficiency of indoor occupants. 

The first drawback of the aforementioned methodology is that it cannot guarantee the indoor working productivity within a reasonable range. The second shortcoming is that the algorithm is not sensitive to the constantly changing environment factors such as temperature, power system condition, DLMP, etc. So the methodology cannot bring the model up to date. To cope with the first problem, we come up with the social cost. the formulation for the social cost comprises two major parts: the utility cost, which is calculated by the end-use energy and the corresponding DLMPs; the cost of work productivity, which is determined by the cost of performance reduction and the number of working personnels \cite{hao2016distribution,jiang2016spat123ial,lu1}. With the increase of buses in distribution power system and the indoor temperature control strategies, the calculation complexity for one game escalates dramatically. Therefore, in this manuscript, we implement markov decision process based multi-agent reinforcement learning to address this problem.

The following sections will briefly review the related research field in this manuscript. Sec.~\ref{sec:introduction021} defines the meaning of "Social Energy" and introduces the computation paradigm in this paper. Sec.~\ref{sec:introduction02} introduces the past and future of distribution locational marginal pricing. Sec.~\ref{sec:introduction03} focuses on the literature of game theory and reinforcement learning in power system. More detailed discussion and reviews can be found in the later chapters. 

\section{Social Energy}\label{sec:introduction021}
The inherent nature of energy, i.e., physicality, sociality and informatization, implies the inevitable and intensive interaction between energy systems and social systems. From this perspective, we define ``social energy" as a complex socio-technical system of energy systems, social systems and the derived artificial virtual systems which characterize the intense inter-system and intra-system interactions. The recent advancement in intelligent technology, including artificial intelligence and machine learning technologies, sensing and communication in Internet of Things technologies, and massive high performance computing and extreme-scale data analytics technologies, enables the possibility of substantial advancement in socio-technical system optimization, scheduling, control and management. We provide a discussion on the nature of energy, and then propose the concept and intention of social energy systems for electrical power. A general methodology of establishing and investigating social energy is proposed, which is based on the ACP approach, i.e., ``artificial systems" (A), ``computational experiments” (C) and ``parallel execution” (P), and parallel system methodology. A case study on the University of Denver (DU) campus grid is provided and studied to demonstrate the social energy concept.

\section{Current Research on Distribution Locational Marginal Pricing}\label{sec:introduction02}

The Smart Grid improves the existing system by accommodating bi-directional flow of both electrical power and real-time communication between consumers and utility operators. Changes to the generation, transmission and delivery infrastructure are supervised, controlled and coordinated by grid operators. Currently, energy efficiency and the emerging of new power loads have further increased the needs for developing new methods for demand side management (DSM). Since the 1980'$s$, DSM has been used as a load shifting tool \cite{gellings1985concept,mohsenian2010optimal,samadi2010optimal} and real time pricing (RTP) is considered as one of the most popular methods that can motivate customers to manage their energy consumption wisely and more efficiently.

In 2010, the University of Denver (DU) facilities spent \$$3.7\,M$ on campus electricity measured from 78 building meters, and 7 buildings have additional demand rate $kW$ ratchet charges. Later in 2011, DU facilities planed to deploy additional methods beyond existing efforts to further lower peak demand, including distributed generation, demand response, proactive heating and cooling, managed load shedding and lighting controls. Driven by the economic goals and regulated by federal laws, DU campus is trying to utilize DSM to reduce peak loads and decrease utility scale in order to control bill demands and cut down $CO_2$ emission. Based on these features, we implement and study the locational marginal pricing (LMP) in the campus power system to generate nodal pricing to help the facilities reduce peak loads, balance power supply and demand, and save on electricity bills. The distribution locational marginal price (DLMP) is modified from LMP to estimate the real-time cost of delivering power to every node in the distribution system and to provide compensation for the renewable energy generation.

Due to the characteristics of LMP, the DLMP based RTP is able to help the electricity market evolve into a more efficient one with less volatility, and the enhanced market efficiency will lead to social welfare for both energy providers and consumers. Smart grid and DLMP will realize DSM and improve elasticity on the demand side, to offer the customer with lower energy cost and to provide the market with increased social welfare \cite{borenstein2009time,schweppe2013spot,lu3}. DLMP has its own advantage through allowing the utility to charge the true cost of electric service to the individual customer rather than mass cross-subsidization \cite{kiefer2011deregulated}. Optimal power flow (OPF) based methods, price area congestion control methods and transaction-based methods \cite{christie2000transmission,lu4} are three techniques to solve congestion management problems. The OPF-based congestion management method is the most precise and efficient methodology and the foundation of the centralized optimization. In \cite{wang2007computational,hao2014Asilomar}, several approaches are demonstrated to deal with the congestion changes in OPF based calculation problem.

DLMP is adopted in this paper for generating real-time prices using the power distribution system model of the DU campus, which is based on the real-world DU utility map. Real world real-time load data is used in the simulation.

\section{Current Research on Demand Side Management, Game Theory and Reinforcement Learning in Power System}\label{sec:introduction03}

Game theory has been implemented in power system for a relatively long time, like the double auction price mechanism \cite{C4MATPOWER}, and it is proven that game theory can realized in abundant scenarios to solve energy related problems. In \cite{SGEVcharging}, the authors present a novel method for energy exchange between electric vehicle (EV) and power grid based on Stackelberg game. However, for academic and commercial buildings, the impact of EV is negligible in terms of the amount of load consumed by EVs. \cite{SGpricingmech} concentrates on the design and implementation of game theory in RTP and load stability for future power grid, and the paper introduces social welfare for maximizing the supplier/consumer profits. Still, the study of the influence of RTP was not included. Some researchers conducted experiments about the relationship between RTP and users' activities \cite{SGjhgPrc1,SGjhgPrc2,jiang2016spat123ial}. But the price mechanisms in those articles,like common flat rate and quadratic cost function, will not fit for future smart grid. Energy scheduling and demand management will benefit smart gird in many aspects such as large scale load shift, mitigate congestion, reduce power system transit stability issues. 

Buildings take up $30\%-45\%$ percentages of global energy consumption \cite{uk2011carbon,jiang2016short}, academic buildings and commercial buildings are labeled as the type of building which consumes the highest power energy within this sector \cite{govregulation1,jiang2016sh123ort,govregulation2,jiang2017sh234ort}. The continuously increasing energy market and $CO_2$ emissions have made the reduction of green house gas and improvement of energy efficiency among the major concerns for energy policies in most countries \cite{whyHVAC1,jiang2015pmu,gu2018multi123}. Therefore, demand side management becomes a very popular and important topic, when people started to pursue higher benefit, e.g. economic profits and social benefits \cite{mohagheghi2010demand, albadi2008summary, vardakas2015survey}. A heating, ventilation and air-conditioning system (HVAC) is universally implemented in large buildings such as academic buildings, shopping centers and commercial buildings \cite{whyHVAC2,zhu2018hierarchical,zhu2017Graphical}. Majority of the research works focus on energy conservation, profit optimization, or pollution elimination problems, and there rarely are works take individual human effect into consideration \cite{brooks2010demand, yusta2007optimal, dong2012distributed, lu2013design, hermanns2013demand, mohsenian2010}. HVAC systems take up the largest energy end usage and impact the cost bill dramatically \cite{whyHVAC1}. And ineffective operations and settings of HVAC systems can lead to remarkably waste of energy, poor indoor air quality and environmental degradation \cite{indoorair,li2017consensus,li2017naps}. Since the ultimate goal of power system development and improvement is to facilitate human life, the effect of human behaviors and their experiences of the services should not be neglected in demand side management and HVAC system scheduling. The biggest obstacle for considering individual human effects in demand side management and scheduling was the extremely high uncertainty and variability of human behavior, making it almost impossible to establish a model for computing. Nowadays, big data and considerably large scale data based modeling techniques can help to find a feasible solution.

In this paper, social cost, which includes the electricity consumption costs and human working efficiency costs, is introduced as an advanced concept to address the importance of both the energy consumption and human experiences in power system management. The optimization of social cost is designated as the objective in this paper to arrange and manage the HVAC system scheduling. Inspired by the methodology in \cite{gamemethod,dai2015werultra,dai2013wefimage,qiu2015computational,qiu2018random}, we propose a game-theoretic solution through formulating the aforementioned problem into a game-theoretic formation. Our proposed approach can solve a finite n-person non co-operative game that the players are the building managers in the same local smart grid, the strategies are the HVAC indoor temperature settings, and the payoffs are the social costs according to various personnels inside those buildings as well as indoor working productivity. It should be noted that we introduce distributional locational marginal pricing (DLMP) to strengthen and reflect the relationship between the plays' payoffs, other player's action and power system situation. To illustrate the proposed methodology and mechanism, we embedded the approach into an interactive real-artificial parallel computing technique. For implementing our methodology and the artificial-real  management/computing system, human efficiency modeling, numerical experiments based smart building modeling, distribution level real-time pricing and social response to the pricing signals are studied. The details of aforementioned techniques will be depicted in chapter~\ref{Chap5}.

Although the game theory can be a good solution for most of the problems, it would still take for a while to solve realistic puzzles and the time would increase exponentially in terms of large size distribution power grid and dozens of control strategies. In terms of our specific optimization objective, there is a need for us to come up with a more advanced algorithm that can solve the proposed objective function faster. To realize the goal, an algorithm needs to be capable of distribution calculation, self learning, and solving discontinuous problem. Hence, multi-agent reinforce learning comes into our searching scope and fits for our optimizing and controlling needs. In this paper, a Markov Decision Processes (MDP) based multi-agent reinforcement learning methodology is implemented to address to optimize the campus social cost.

\section{Architecture of the Paper}
In Chapter 1, the literature of current research are studied, which provides the motivation, rationale and background for the paper. In Chapter 2, the computational paradigm of the research is demonstrate. And social energy is also introduced in this chapter. To reflect the cost of energy, Chapter 3 mainly introduces the distribution locational pricing, and the implementation of the DLMP. University of Denver campus power system is used to demonstrate. Chapter 4 illustrates the preliminary methodology that is used to investigate our objective function. Chapter 5 focuses on the advanced methodology that we implement in our research to improve the computational efficiency. The related numerical experiments are conducted in the University of Denver power system, which is a $57$ buses distribution power system. Chapter 6 concludes the report.

\chapter{Chapter 2}\label{Chap2}
\section{Introduction and Motivation for Parallel Computing and Social Energy}
Energy has always been a key element in the development and operation of a society. It is the backbone which supports the prosperity of a modern society. It is the hand that pushes the society to move forward, meanwhile, it is also one of the major limitations and barrier which hinders the pace of social development \cite{hao2017SocialEn}. 

\subsection{Grand Social Challenges in Energy and Power System}
The trend of electrification forced by the Second Industrial Revolution eventually triggered the explosion of the demand for energy supply, and set global social development on the basis of fossil fuels. Based on the statistics provided by The World Bank, fossil fuels, including coal, petroleum, and natural gas, have accounted for over 80 percent of world's energy consumption ever since 1990. Although the development of the energy industry constantly propels the development of the entire society, problems, as many can see, have emerged \cite{Brown2015,NY09,Werner2015}. The first one arising is the depletion of fossil fuel resources, which was foreseen even at the early stage of their large-scale utilization. Another most obvious problem, which catches global attentions, is the environmental impact and the climate change. Pollutants, including carbon monoxide, oxides of nitrogen, sulfur oxides, hydrocarbons and particulate matters, which are released from fuel combustion, are contaminating the air we breath everyday and encroaching on human health. Uncontrolled oil spills, coal mining washing, and acid rain are polluting the water and damaging the aquatic ecosystem. And strip mining and some closed power plants have left large area of land fallow and wasted. If environmental pollution only affects certain areas, where fossil fuels are exploited, transported and processed, climate change, mainly global warming, affect the life of all human beings. Ever since the First Industrial Revolution, increased dependence on fossil fuels has resulted in a huge amount of green house gas emission and has dramatically accelerated the global temperature rise. Melting glaciers, rising sea levels, swallowed continent, reduced food production, species extinction, and ecosystem collapse can all come to reality if the temperature rise continues to follow the current pattern. Other problems, such as political or security problems also arose among nations due to the uneven distribution of the fossil fuel sources.

Being aware of the benefits as well as the side effects brought by conventional energy resources, societies worldwide have already started to take actions, i.e., reducing their dependence on conventional energy and transiting to alternative energy resources. Nuclear energy was once thought to be the substitute \cite{IAEA2014}. Although it produced $11$\% of the world's electricity in 2014 \cite{Mycle2014}, its contribution is declining due to the climbing cost and, most importantly, due to its potential destructive danger, which was exposed to the world through the Chernobyl and Fukushima accidents. Clean renewable energy is receiving greater expectations \cite{Brown2015}, which includes hydro power, solar power and wind power, which are inexhaustible and accessible almost anywhere in the world. Many large-scale solar and wind power plants have been established worldwide. Meanwhile, solar and wind energy have been accepted at the commercial and personal consumer level, e.g., small-scale renewable energy production for commercial buildings and residential housings. The above progress, together with the emergence of a new energy utilization form, electrical vehicles, have changed the role of the passive end consumers to a more active one as prosumers (producer-consumer) \cite{prosumer}.

The diversity of energy sources and involvement of social entities are changing the structure of the power system by incorporating distributed generation and storage capabilities into the conventional centralized operation mode, making it much more complicated and much difficult to operate \cite{Bian2014, Wigan2014}. Efficiency and security are among the most severe concerns. How can an energy system incorporate different energy forms? How to take the advantages of different energy forms in order to enhance efficiency while eliminating waste and side effects? And how could an energy system be operated to maintain stable performance to ensure secure generation and transmission in case of any type of disturbances from both inside and outside of the system. The quest for an intelligent system now is urgent than any other time in the history. The system should be able to dig, collect, process, digest and utilize the tremendous information flowing in and between every parts and procedures for the purpose to monitor, manage or even provide advices to itself.

\subsubsection{Physicality of Energy}
All the processes of energy production and consumption take place in the physical space. The major energy resources encompass fossil fuels, e.g., coal, petroleum, and natural gas, alternative and renewable resources, e.g., hydro, solar, wind, biomass and geothermal, and nuclear energy. They are used to provide cooling, heating, illumination, mechanical power and electricity. The devices and systems involved in energy transformation and utilization include boiler, steam engine, steam turbine, electric generator, electric motor and many advanced devices and systems which are composed of various electricity consuming equipment. Furthermore, the rising of distributed energy sources, energy storage, combined cooling, heat and power (CCHP), and electric vehicles have further enhanced the diversity and complexity of the devices and systems working in the energy production and consumption process.

\subsubsection{Sociality of Energy}
Ultimately, energy is produced to serve the human society. Thus, inevitably a label of sociality is attached to it. Sociality of energy is presented and stated in the following three aspects \cite{hao2017osti}.
\paragraph{Direct Involvement of Human Beings during Energy Production}
Human beings directly participate in every procedure of energy production and consumption, i.e., planning, designing, constructing, operating and maintaining energy systems. Different knowledge background, proficiency levels, subjective consciousness and even emotional status of the participants in these procedures might affect the system in different ways. For example, different operators may result in different efficiency levels even when operating the same boiler under the same condition for thermal power generation. Per statistics, the fluctuation in efficiency can at least reach 0.5\%. When the boiler is working at an unrated state, the effect is even more prominent. Therefore, the energy production process can reflect one aspect of the sociality of energy.
\paragraph{Sociality Reflected in Load Characteristics}
Influenced by the applications of many demand side management programs, consumers would probably change the schedules of their daily activities according to real time utility prices to pursue lower costs. And this would definitely lead to load change together with human's activities, habits and mental states. Besides, since an increasing portion of population are choosing electric vehicles as their means of transportation, load becomes to shift not only in the time domain but also in the spatial domain in response to people's needs for traveling, people's decisions, electricity prices and to the traffic conditions. What's more, since the quality and capacity of energy storage devices have been largely enhanced, people now can store energy at off-peak time, move it to complement peak time usage and they even have the option to trade with the grid. And this new possibility can help enhance power system security, improve efficiency and save costs. Hence, the load characteristics can reflect another aspect of the sociality of energy. possibility can help enhance power system security, improve efficiency and save costs. Hence, the load characteristics can reflect another aspect of the sociality of energy. Therefore, the need for design a novel computation algorithm, which can taking care of the competition between prosumers and time-varying load profile, is urgent for future energy and power system. 
\paragraph{Sociality Reflected in Load Characteristics}
The planning of an energy system is constrained by all kinds of considerations, i.e., the energy sources, environmental endurance, economic conditions and the population. As more issues have emerged due to the utilization of conventional energy, optimized control and management of multi-source energy production becomes even more urgent. Many countries have established governmental mechanisms to support and encourage the development of environment protection programs. As a result, the penetration rate of renewable energy in energy production is rising rapidly, which creates new challenges to the current power system for maintaining stability and security, and requests all the energy generation units to work under varying operating conditions. How to operate the thermal generating units to ensure efficiency while reducing pollution? How to incorporate the requirements and expectations from the government and society, such as limitations on emission and required rate of renewable installations? These questions would be vital for the future power system to answer for achieving flexible energy system operation. Thus, energy system operation, which is restricted by policies and regulations, can reflect the last aspect of sociality of energy.

\subsubsection{Informatization of Energy}
Energy flow itself provides huge amount of information. Although the information is easy to acquire, yet few systematic analyses have been conducted to better utilize the value embedded in them. Information is power, as long as it is employed reasonably and efficiently. To assist the current energy system, the construction of an information system is necessary. The information system could offer an overview of the current situation of energy use by analyzing historical data, reconstruct the frame of each process, and provide optimized plan for future use aiming at enhancing efficiency and slowing down energy depletion. In any energy system process, an information flow is coupled with the energy flow. The information generated in sensing, computing, monitoring, scheduling, dispatching, and control and management directly affects and dominates the energy flow, and leaves a sign of informatization on the energy system. Besides, the dominating status of information in the energy system also enables it to receive information from the society and to be heavily influenced by human's thoughts and judgements. Thus, the informatization of energy system also reflects the property of sociality in energy.

To respond to the grand challenges and based on the inherent properties of energy, we propose the definition of ``social energy", which is enabled by the recent advancement in intelligent technology, including artificial intelligence, and machine learning, sensing and communication in Internet of Things technologies, and massive high performance computing and extreme-scale data analytics technologies. In the following, we first provide a proposal on the general methodology of establishing and investigating social energy, which is based on the ACP approach, i.e., ``artificial systems" (A), ``computational experiments” (C) and ``parallel execution” (P), and parallel system methodology, and then propose the social energy systems for electrical power. A case study on the University of Denver (DU) campus grid is provided and studied to demonstrate the social energy concept. In the concluding remarks, we discuss the technical pathway, in both social and nature sciences, to social energy, and our vision on its future.

\section{General Methodology in Establishing Social Energy}\label{sec:approach}

\subsection{The ACP Approach}

\begin{figure}[!h]
	\centering
	\includegraphics[width=5.7in]{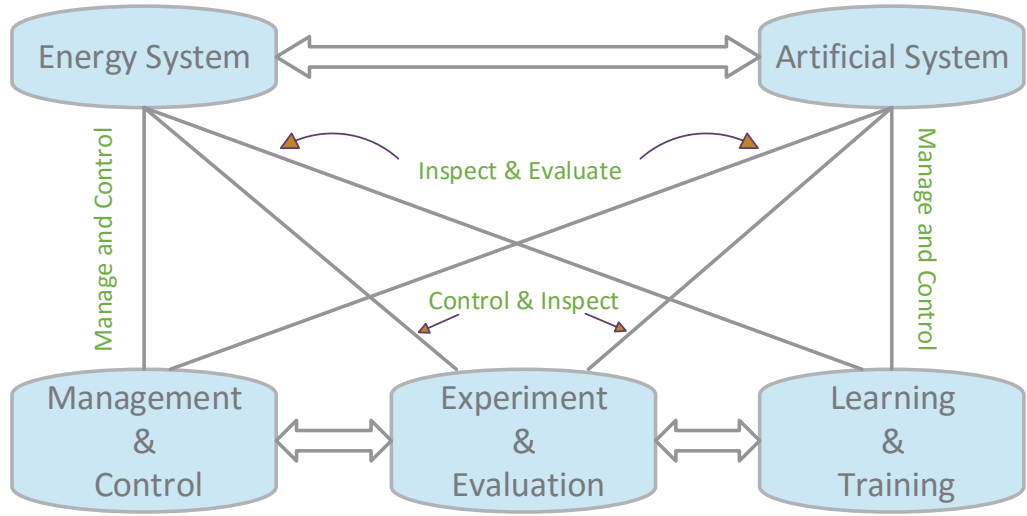}
	\caption{The ACP approach.}
	\label{fig:ParaEx}
\end{figure}

The ACP approach consists of ``artificial systems" (A), ``computational experiments" (C) and ``parallel execution" (P) \cite{5wang2007toward,wang2013intelligent}.
The ``artificial systems" serve for building complex models based on the data and information collected from the real physical world by employing data-driven approaches and semantic modeling methods. They apply Merton's laws to construct a feedback exchange mechanism between information and behavior. The ``computational experiments" aim at accurate data analytics. Since numerical representations for human society activities can hardly be extracted for quantitative analysis, social computing methods must be applied. Social computing methods \cite{5wang2004social} such as deep computing, group breadth computing and historical experience computing can help obtain the results from different modes of the virtual artificial systems. The social computing methods must be rooted in the human society, utilizing artificial intelligence instead of traditional computation methods to model the society. And the ``parallel execution" targets at innovative decision-making. The artificial energy system and physical energy system form a pair of parallel energy system, constructing a new feedback control mechanism based on virtual-physical interactions.

Recent development of new IT technology, such as deep learning, Internet of things, and cloud computing, lays a technical foundation for the realization of the ACP approach. The core philosophy of ACP is to interpret and transfer what is virtual in the complex CPSS to quantifiable, computable and processable processes, which turns the virtual artificial space into another space to solve complexity problems. The ACP approach targets at establishing a ``virtual vs. real" parallel system of a complex system, and solving complex system problems through real-time computing and interacting between the two. Generally speaking, the ACP approach contains three majoy steps: 1) Modeling complex systems using artificial systems; 2) Evaluating complex systems using computational experiments; and 3) Realizing effective control and management over the complex system through interacting the real physical system with its virtual artificial system.

The virtual artificial space and the real physical space form the ``complex space" for solving complex system problems. The most recent development of new IT technology, such as high-performance and cloud computing, deep learning, and Internet of Things, provides a technical foundation for the ACP approach. In essence, the ACP approach aims to establish both the ``virtual" and ``real" parallel systems of a complex system, and both of them are utilized to solve complex system problems through quantifiable and implementable real-time computing and interacting. In summary, the ACP approach is consist of three major steps. 1) Using artificial systems to model complex systems; 2) Using computational experiments to evaluate complex system; and 3) Interacting the real physical system with the virtual artificial system, and through the virtual-real system interaction, realizing effective control and management over the complex system.

In a sense, the ACP approach solves the ``paradox of scientific methods" in complex system's ``scientific solutions". In most complex systems, due to high complexity, conducting experiments is infeasible. In most cases, one has to rely on the results, i.e., the output from the complex system, for evaluating the solution. However, ``scientific solutions" need to satisfy two conditions: triable and repeatable. In a complex system that involves human and societies, not being triable is a major issue. This is due to multi-faceted of reasons of prohibitive costs, legal restriction, ethics, and most importantly, impossible to have unchanged experiment conditions. The above inevitably results in the ``paradox of scientific methods" in complex system's ``scientific solutions". Thus, the ACP pursues the sub-optimal approach of ``computational experiments". The computational experiments substitute the simulations for physical systems when simulations are not feasible. In this way, the process of solving complex system issues becomes controllable, observable and repeatable, and thus the solution also becomes triable and repeatable, which meets the basic requirements of scientific methods.

\subsection{Parallel Control System}

Based on the ACP approach, the parallel intelligence can be defined as one form of intelligence that is generated from the interactions and executions between real and virtual systems \cite{wang2013parallel}. The parallel intelligence is characterized by being data-driven, artificial systems based modeling, and computational experiments
based system behavior analytics and evaluation. The core philosophy of parallel intelligence is, for a complex system, constructing a parallel system which is consisting of real physical systems and artificial systems. The final goal of parallel intelligence is to make decisions to drive the real system so that it tends towards the virtual system. In this way, the complex system problems are simplified utilizing the virtual artificial system, and the management and control of the complex system are achieved. Fig.~\ref{fig:ParaEx} demonstrates a framework of a parallel system. In this framework, parallel intelligence can be used in three operation modes, i.e., 1) Learning and training: the parallel intelligence is used for establishing the virtual artificial system. In this mode, the artificial system might be very different from the real physical system, and not much interaction is required; 2) Experiment and evaluation: the parallel intelligence is used for generating and conducting computational experiments in this mode, for testing and evaluating various system scenarios and solutions. In this mode, the artificial system and the real system interact to determine the performance of a proposed policy; 3) Control and management: parallel execution plays a major role in this operation mode. The virtual artificial system and real physical system interact with each other in parallel and in real-time, and thus achieve control and management of the complex system. We would like to point out that one single physical system can interact with multiple virtual artificial systems. For example, corresponding to different demands from different applications, one physical system can interact simultaneously or in a time-sharing manner with the data visualization artificial system, ideal artificial system, experimental artificial system, contingency artificial system, optimization artificial system, evaluation artificial system, training artificial system, learning artificial system, etc.

\begin{figure}[htp]
	\centering
	\includegraphics[width=5.7in]{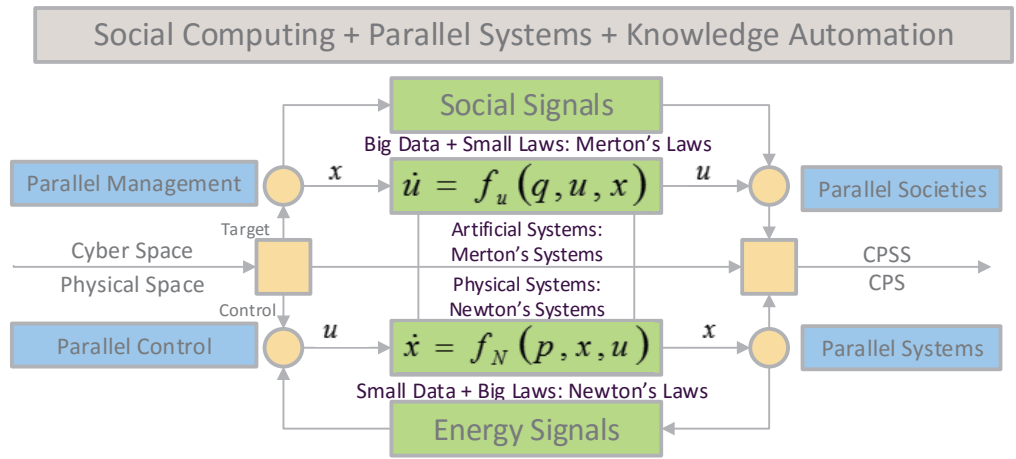}
	\caption{The parallel system approach for real-virtual interactive system control and management.}
	\label{fig:ParaFrame}
\end{figure}

The ACP approach and parallel system methodology have been applied to several major social challenges, including intelligent transportation systems, social elderly healthcare systems, mass production system management, and social computing and management for security, and achieved remarkable success. The example of Intelligent Transportation Systems (ITS) in Urban Areas is demonstrated as following. The major challenges in ITS lies in two folds: \textit{1.)} Transportation system data are very difficult to obtain neither from the related government departments nor from field experiments, resulting in infeasibility in large-scale and close-to-reality study of transportation systems; \textit{2)} Current intelligent transportation systems rely on data from past events for decision-making for future events, lacking a mechanism of investigating the actual causes of ITS events. The ACP approach is employed to study this socio-technical complex system, utilizing the parallel ``Artificial Transportation Systems" (ATS) for modeling, control and management, which is able to bypass the two difficulties stated above \cite{Wan13,Wan08,Wan07}. Utilizing the ACP approach, the social factors, such as population distribution, individual human behaviors, climate and public emergencies, are taken into considerations in the ITS system. Together with the cloud-based computing and IoT technology, the traffic conditions can be forecast. Meanwhile, the ACP approach can also provide suggestions on how to improve traffic conditions through control and management before the actual traffic congestion takes place.

The success of the ACP approach and parallel system methodology provides a feasible and promising technical pathway for studying the proposed social energy system, which is illustrated in the next section.

\section{Definition and Unique Characteristics of Social Energy}
\subsection{The Definition of Social Energy}

Utilizing the concepts and methodology introduced in the previous sections, we provide the definition of the proposed social energy as following.

A social energy system is a complex of physical energy systems, physical social systems, and the artificial virtual systems derived from the physical systems. The artificial virtual systems are derived with certain purposes that concern the joint operation of the socio-technical systems. Utilizing the multifaceted data collected from the socio-technical systems, through sufficient interacting and massive computing, knowledge automation of the systems is gained, and intelligence in system control and management is generated. The knowledge and intelligence in turn are applied in the social energy system, achieving a truly automated and intelligent joint socio-technical system design and management.

We propose to use the general methodology of ACP approach and parallel system in establishing the proposed social energy system, and its unique characteristics are explained as the following.

\subsection{Unique Characteristics of Social Energy}

\subsubsection{Artificial Systems}

In the era of Industrial and Energy 3.0, the power system and information system are designed separately, bringing about a lack of interaction between the two, and, to some degree, resulting in the lack of utilization of the informatization nature of energy. Coming into the era of Energy 4.0, the concept and frame of a ``Cyber physical power system'' have been proposed, focusing on the amalgamation and cooperation of the power and information systems, especially the influence of the information system upon the power system. However, the specific theories and techniques needed to build such a system are not yet mature.

Most of the studies on the physical energy systems focuse mainly on the energy flow in the processes of energy transformation, combined utilization of multiple energy sources, cascaded utilization of energy, environmental friendly substitutes for conventional energy sources, etc.

The design and plan for conventional power system control and management schemes usually only consider the physicality of the energy flow, and ignore, most of the time, the sociality and informatization properties of energy. With the rising of intelligent energy transformation, transmission and consumption, the energy system is required to incorporate the intermittent renewable sources, to catch up with rapid demand changes, and to actively meet the requirements of saving energy and lowering emission, which requires in-depth incorporation of the social system information, i.e., individual, organizational and social behavior information which contains complexity and uncertainty. To explore how these social elements affect the energy system, and further to assist improving the socio-technical system's design, operation and maintenance, one should understand and combine several disciplines, such as sociology, management, economics, anthropology, and praxeology.

The practice for developing the aforementioned interaction and cooperation mechanisms among the physical, social and informative properties have not been launched yet, and they cannot be realized in the conventional power system simulation models, which usually only take minimum social information into considerations, such as power demand. In addition, the cooperation strategy among the three components of the cyber-physical-social system should be adaptively updated in accordance with the real-time working conditions, in order to provide intelligent decisions. Unlike the traditional deterministic or probability-based simulation models, most of the changing conditions cannot be forecast, which means the structure of the system model should be time-varying.

Conventional power system simulation models consider and utilize only the load patterns, while the social elements, e.g., incentive mechanisms, people's consumption habits or people's decision-making patterns, involve knowledge from multiple disciplines, which fundamentally denies the feasibility of the conventional simulation methods to build the social models. Besides, the power system is a non-linear system, and most of the simulation models are built and serves around a certain nod in the grid, that they can hardly reflect the overall working condition of the entire grid, resulting in a relatively conservative operation style. What's more, due to the fast development of the ''Internet of energy'', multi-source energy combination and demand side management become inevitable, which further elevates the degree of system non-linearity. Therefore, the complex working conditions of the power system cannot be reliably and effectively realized by the existing simulation techniques.

Revolutionary theories and techniques are in urgent demand to analyze, manage and control such a complex system. To response to the call, the concept of artificial energy systems based on the idea of parallel systems is deemed by us as a suitable methodology for social energy. The uncertainties existing in the cyber-physical-social system and the complexity of the human society intrinsically determine that this system would be extremely complicated. Hence, to transit from Newton's Laws to Merton's Laws, from control to guidance, the establishment of artificial systems is in an urgent need.

The virtual artificial energy system will be established in the ``cyber space" based on semantic \cite{5wang2015}, data driven or other modeling techniques. It should work parallel to the physical energy system, such that the two systems interact and exchange feedback between each other. The artificial system should be able to reflect the real-time working conditions of the physical system, and on the other hand, it needs to conduct parallel computations to provide optimized solutions and advices for better performance of the physical system. Taking this as a precedent, it is believed that introducing the virtual artificial system into the complex energy control system will definitely start a revolutionary new era in the energy industry.

\subsubsection{Knowledge Automation Through Interacting and Computing}
A social energy system usually contains highly complex physical and informative processes. The complexity of the social energy socio-technical system exceeds by far the complexity that the industrial automation systems can handle, which means conventional industrial automation approaches, including manual control, single loop, multiple loop, distributed control system (DCS), manufacturing execution system (MES), enterprise resource planning (ERP), etc., cannot satisfy the requirement of social energy any more.

The artificial system derived from the social and energy systems will play a central role in enabling the processing of tremendous data and information from the energy and social systems. The physical system data and information with features of uncertainty, redundancy, and inconsistency, which decide that human intelligence alone has no ability in processing and analyzing the data and information. Therefore, it is necessary to initiate the system of knowledge automation \cite{5wang2015software}, employing data-driven methods, multi-agent system and other artificial intelligence techniques to liberate human intelligence and to achieve the desired outcome of the artificial system. The approach for realizing data automation is introduced in Section~\ref{sec:approach} as the ACP approach and parallel system methodology.

\subsubsection{Socio-Technical Feedback Mechanism and Gamification of Reality}

The direct outcome of the intensive interaction among social, technical and artificial systems is the feedback mechanism that directly or indirectly leads to the desired system goals. We note that the physical socio-technical system can interact with multiple virtual artificial systems, e.g., consumer behavior artificial system, organizational behavior artificial system, contingency artificial system, scheduling and optimization artificial system, real-time management artificial system, etc. Each of the artificial system corresponds to different applications. And the knowledge automated through interacting and computing generates feedback signals which are used for modeling and shaping the real physical social and energy systems. The feedback may range from direct control of technical system, indirect manipulation of distributed socio-technical agents, prescriptive influence on social and economic entities, publicity for maximum information dissemination or even psychological hints on individual behaviors. In \cite{Wer12}, the effective and efficient feedback is characterized as the core feature of gaming from philosophical perspective. In the social energy socio-technical system, strong, effective and efficient feedback mechanism provides the possibility of the gamification of reality, i.e., turning the reality into a ``game", where the physical systems are shaped and pushed toward the desired ones much more rapidly. With such strong feedback mechanism, the socio-technical system is expected to possess advanced closed-loop operation capability with stability, rapid system convergence and agility in system adaptation.

The era of energy 5.0 will be an era of big data. The sensing and surveillance data from physical systems, pertinent data from information control, and social activities and policies all will be the data sources. There will be no concrete models or references for the conventional simulation techniques to imitate, thus the conventional simulation and control modes can hardly handle the challenges from the era of big data. A data-driven parallel control mechanism utilizing the theories, methods and techniques of knowledge automation will be the core in the establishment of the artificial system. The focuses and key issues of the CPSS are illustrated as the following.
\paragraph{Scientific problems} While traditional computations and physical models work independently, the CPSS requires a unified modeling theory to realize dynamic interactions among computational, physical and social processes, achieve time-space consistency, cope with uncertainty problems, and eventually to accomplish the parallel operation of the physical and artificial systems.
\paragraph{Technical problems} Corresponding to the above scientific problems, the technical focuses should be put on the development of new scheduling, design, analysing and experiment tools which can utilize social computing, parallel execution or other strategies in order to reflect the activities of interactions and evolutions.
\paragraph{Engineering problems} The engineering problems mainly encompass system construction, design, integration, maneuverability \textit{etc.} Issues like reasonable time managing of the physical system and concurrency of the physical and virtual artificial systems are also of importance.

\begin{figure*}
	\centering
	\includegraphics[width=5.7in]{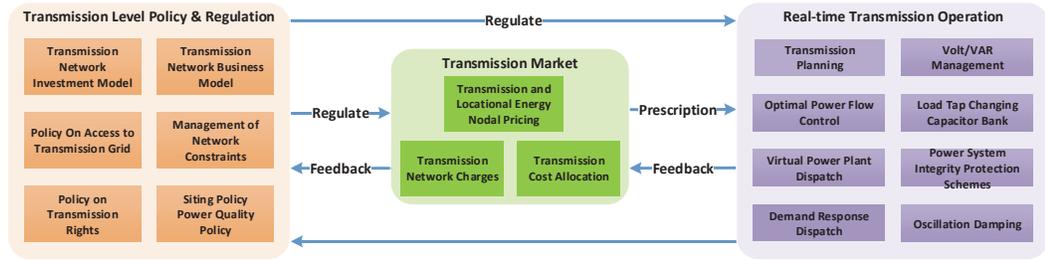}
	\caption{The socio-technical power system architecture at the bulk generation and transmission level.}
	\label{fig:Trans}
\end{figure*}

\section{Architecture of Social Energy for Electrical Power System}

In this section, we provide an initial investigation on the socio-technical system architectures for electrical power systems at different levels for demonstrating the interactions among electrical energy systems, social systems and their resulting artificial systems.

\subsection{Bulk Generation and Transmission Level Architecture}

As the first chain in the electricity delivery, bulk electricity power generation and transmission networks aim to generate and transmit electrical power that balances power demand across large geographical areas. Therefore, the most important issue faced by the transmission network is how to efficiently dispatch electricity from generation to load, which includes the efficient operation of existing transmission facilities and the expansion of the network to meet increased demand.

At the stages of transmission grid construction and expansion, regulatory forces carry the responsibility to decide the network investment, business models and access policies to set the basic tone of grid development \cite{Regul, Elforsk, Grid2030, christie2000}. In the operation of transmission grid, ``economic dispatch" means that the least expensive energy is transmitted to meet power demand while ensuring the security and stability of the power system. In an ideal scenario, simple employment of generations with least expensive operation costs would be the best solution to achieve the goal. However, network effects, which encompass network losses, grid-imposed constraints, and quality of service, are the main factors that influence the operation of the transmission grid. Regulatory, economic and technical efforts are designed to cope with the impacts brought by the network effects.

In terms of quality of service, although most of the quality issues occurring in the demand side are due to power distribution level failure, the less common issues caused by transmission failure generally bring severe consequences to the entire power system \cite{Dobson2007}. This is determined by the characteristics of electricity flow, which obeys the Kirchoff's laws, that failure in one working transmission line immediately triggers the Domino's effect and the failure is spread across the grid to induce black-out. To prevent this kind of catastrophe, power quality related policies are established, and in response a series of technical measures are taken to preserve the stability of the grid. Techniques such as Volt/VAR control and load tap changing capacitor bank are developed to maintain reasonable active and reactive power amount in the grid to ensure proper power flow and voltage levels. And power system integrity protection,  oscillation damping and other related techniques intend to minimize the impacts and protect the grid from disturbances.

Network losses \cite{glover2011power, wood2012power}, which consists of mostly ohmic losses and losses due to corona effects, add the influence of location to the ideal scenario of energy dispatch, that prices in the transmission grid are different from node to node. Grid-imposed constraints, which include the constraints resulted from the physical limitations of transmission lines and facilities, e.g., current flow limits and the constraints established by the regulators considering the security issues of grid operation, e.g., capacity and voltage level, further emphasize the locational effects and complicate the process of energy dispatch for pursuing the most economical way.

Transmission level nodal prices (equivalent to locational marginal prices) \cite{schweppe1988, borenstein2005time, crew1995theory} are the marginal prices at each node, counting the impacts of losses and constraints, and provide the locational pricing signals. The nodal prices play a truly important role in the transmission market. It enables the operators to remunerate utilizing the price differences. It provides the information for guiding transmission planning, optimal power flow control, virtual power plant dispatch and demand response dispatch, and therefore it is meant to achieve the ``efficient energy dispatch". Also for the regulators and policy makers, the locational signals can provide useful advices on siting policies.

The nodal pricing mechanism cannot cover all the power transmission investors' costs, in addition transmission charges must be issued to the beneficiaries of the grid. And an effective and efficient cost allocation mechanism should allocate the cost for initiation, operation, maintenance and expansion to all the beneficiaries properly, that everyone's profits are ensured and the economic viability of the system is protected.

\subsection{Distributed Generation and Distribution Level Architecture}\label{sec:distri}
\begin{figure*}
	\centering
	\includegraphics[width=5.7in]{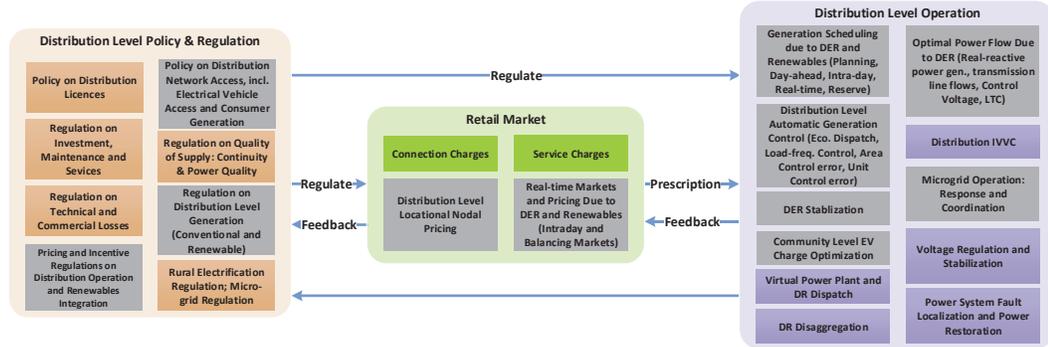}
	\caption{The socio-technical power system architecture at the distributed generation and distribution level.}
	\label{fig:Distr}
\end{figure*}

Compared with the bulk power generation and electricity transmission grid, the distribution grid is a much more intricate system \cite{mei2011power}. It connects to the transmission grid, steps down the transmission voltage to meet consumption's requirements and deals directly with the various end users, fulfilling the last step in delivery of energy from generation to consumption. The rise of distributed generation (DG) today \cite{borbely2001, ackermann2001, barker2000} allows smaller scale electricity generations to be connected to the distribution grid. The functions of and responsibilities carried by the distribution grids determine the complexity and massive quantities of their components and operation mechanisms, which implies that more regulatory, technical and economic efforts are required to maintain its viability and functionality.

The most basic duty of the distribution grid is to deliver electricity to the end users within its service area for supporting resident living and society development. The idea sounds simple, however, it requests the fulfilment of a series of complicated tasks. In the initial stage, the distributors/operators of the distribution grids must plan, design and build the capacity and structure of their distribution network by analyzing and evaluating the current demand and possible increase in demand in the future. During the life time of services, distributors need to properly operate the grid and develop techniques to ensure continuity of provision of high quality electricity. Non-discriminatory acceptance of new-coming consumers should be ensured and their connection to the grid is carried out by the distributors. For long-term steady and secure services, it is also the responsibility of the distributors in conducting regular and effective maintenance of the system.

All the tasks mentioned above require huge investment. Although it is the distributors' duty to maintain a complex distribution network which can deliver high-quality electrical power to the end users, the distributors' ultimate goal is always to profit as much as possible. Following this logic, the distributors may want to reduce the costs invested, as a result, conflicts between the consumers' and the distributors' benefits emerge. Distributors may intend to deny the access of new consumers to the existing distribution grid, for avoiding the costs for extra capacity planned in the initial grid construction, and the costs for new connection nodes. In terms of quality of service, for the distributors, higher quality implies higher costs, while consumers always prefer high-quality power. Rural or remote area electrification is another example, which are costly and low-profit projects to distributors, but they are imperative due to governments' regulation and beneficial for the residents in those areas.

The most severe conflicts today would probably arise around the proliferation of distributed generations (DG), distributed storage and electric vehicles (EV) \cite{clement2010,pepermans2005,quezada2006,cheng2014electrified,cheng2015d2d,zhang16flex,zhang16energy,Cheng16con,globalcom}. Considering the environmental friendly nature of renewable generations, the increased energy efficiency using co-generation of combined heat and power (CHP), and the governmental promotion and incentive policies to support distributed generations, more consumers may be willing to deploy DG. Despite the advantages DG can bring, it would impact the existing distribution network and the distributors' benefits in many ways. Since the current distribution grids are designed for unidirectional energy flow (from transmission to distribution to end users), connections of new DG facilities request investment on new technologies and constructions. And the reverse and extra power flow from users back to the grid can impact the security, stability and quality of the energy delivery, and entails the distributors to restructure their operation modes.


Policy makers and regulators influence the entire system by direct forces, i.e. direct regulations and indirect forces, i.e. indirect prescriptions through retail market. Regulators render qualifications and establish obligations, such as service duration, area of service, regular maintenance requirements, etc., through distribution licensing. And by means of incentive-based regulations, distributors are always encouraged and awarded to pay more efforts achieving certain reference level requirements, and are penalized for not accomplishing required works. Techniques such as distribution integrated Volt/VAR control (IVVC), voltage regulation and stabilization, power system fault localization and power restoration are the responses to those regulations for improving service quality, maintaining system security and stability, and reducing system energy losses. Reasonable prices are regulated by evaluating the market and the cost reports submitted by the distributors, in order to protect both consumers' and distributors' benefits. Feedbacks from the retail market and distributors in turn help the regulators to renew and adjust prices and those reference levels to adapt to changes and keep up with the trend of system development.

Although the future of DG seems promising, the scale of DG installation today has not reached the level to severely impact the distribution grids. Whereas, since DG's deployment will be an inevitable trend, corresponding to regulatory complements, market pricing mechanisms and technological developments should be initiated to cope with the challenges that DG will bring. Specific and more detailed regulations would be carried out to formalize the connection of DGs and EVs to the distribution grid, and to regulate the generation and operation of DGs. Prices would be set at a much smaller time-scale \cite{Itistime,li2014distribution} e.g.， day-ahead, intra-day or real-time, to incorporate the characteristics of DGs, e.g.， intermittence of renewable energy generation. And the charges for consumer side DG connections and maintenance should be set properly. For the distributors, approaches such as generation scheduling due to DG, automatic DG control, DG stabilization, optimal power flow due to DG and EV charge optimization are of great value and importance to optimize grid performance by affiliating DGs.

\subsection{Consumer Level Architecture}\label{sec:consumer}

\begin{figure*}
	\centering
	\includegraphics[width=5.7in]{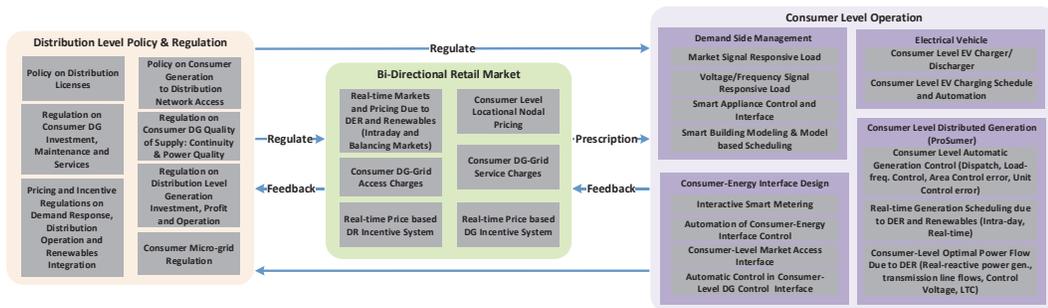}
	\caption{The socio-technical power system architecture at the consumer level.}
	\label{fig:Consum}
\end{figure*}

In traditional power system, consumers have been treated as solely passive end users. Nowadays, the rapidly increasing number of distributed generation installations at the consumer level and the growing interests in electrical vehicles enable the consumers to be involved as active players in the power system. Furthermore, research from various fields, such as sociology, anthropology, economics and engineering, has realized that consumer behavior has great influences on the working status of the system, that if the consumers' energy consumption activities can be correctly guided, both the system and the consumers themselves can obtain pronounced benefits. Therefore, it is necessary and urgent to establish an interactive framework, like in all the other levels in the power system, which consists of the corresponding regulatory and market mechanisms as well as technical supports and operational tactics, to better direct and manage the energy related activities of consumers.

Unlike the transmission and distribution levels, where machines take most of the works under exact and unified instructions of the operator, in the consumer level, almost all the operations are determined and fulfilled by different individuals with different characteristics. Their decisions and activities possess high uncertainty. This uncertainty together with the uncertainty of distributed energy sources make the system control and management in this level very difficult. Therefore, ``intelligent" techniques and automation plays extremely important role at this level to help simplify and unify the control and management processes \cite{farhangi2010path,gungor2011smart,di2012event}.

We define a new concept of ``Consumer-Energy Interfaces" (CEI) at the consumer level, which will be the intermedia empowering the interactions between consumers and the network. Interactive smart meters allow both sides to exchange information about the working conditions of each other; consumer-level market access interface enables the bidirectional trades of energy; and automatic consumer energy control interface and automatic consumer level DG control interface can ensure the safe and efficient energy flow. As the entrance for consumers to participate in the grid, significance of the interface design is conceivable. To ensure the entrance for consumers to participate in the grid, but at the same time it should be able to provide all the important information for its users to make proper decisions. However, what information is considered important to a particular user? And how simple the interface should be? Many more questions need to be well studied and answered by the designers. And according to the answers, regulators should establish pertinent standards to regulate the use of the interfaces, so that consumers' security and privacy should be sufficiently protected.

Given the capability to interact and exchange information with the market and the energy network, the efforts that can be made to maximize the benefits and minimize the adverse impacts of consumers' involvement (act as Pro-Sumers) \cite{prosumer} are discussed in the following. At first, demand side management (DSM) strategies \cite{fahrioglu2001using,DSM2010autonomous,DSM2011demand} should be improved and widely applied to relieve the burden of peak time generation and transmission, and from a long-standing point of view to save investment in all the facilities to meet peak demands. Market signals like real-time prices and consumer level locational nodal prices are inevitable elements in DSM, which reflect the real-time working status of participating entities, e.g., demand changes in consumer side and available grid capacity. Based on the pricing signals and the exchanged parameters provided by smart meters, and supported by smart appliance and load control interface, demand management programs such as market signal responsive load and voltage/frequency signal responsive load can be realized. While these measures are based on real-time control and management, smart building modeling can provide advises on scheduling \emph{ex ante}. When the consumers play the role as energy producers utilizing local DGs, as mentioned in the previous subsection (Sec.~\ref{sec:distri}), it brings not only benefits but also impacts to the grid. Since the relative efforts to be made have been narrated in Sec.~\ref{sec:distri}, although not limited to those have been discussed, tautology is avoided here.

To achieve the ``healthy" energy flow involving prosumers, one can predict that huge information flow is also transmitted within the grid. This phenomenon may initiate the unique and non-negligible issue of security and privacy, which did not rise in any other levels. Regulatory institutes should carry part of the responsibility to take care of this issue, and technical solutions, e.g., information security, should be advanced for better protection of the consumer and the grid exposed to the information networks.
\chapter{Chapter 3}\label{Chap3}
\section{Distribution Locational Marginal Real Time Pricing}\label{sec:WAMs}
The first step to build the artificial system and the computing paradigm introduced in chapter~\ref{Chap2} is to establish a real time pricing mechanism. Our research, presented in this manuscript , mainly focuses on distribution power grid. Hence, to achieve the ACP approach and parallel computing scheme in chapter~\ref{Chap2}, and demonstrate the architectures in Sec.~\ref{sec:distri} and Sec.~\ref{sec:consumer}, this chapter introduces the locational marginal real time pricing mechanism in distribution level that can reflect the nodal prices dynamically. 
\subsection{Extraction of Network Topology Based on DU Campus Utility Map}
\begin{figure}[!t]
	\begin{center}
		\subfigure[]{ \label{fig:DULoopnetwork}
			\resizebox{3.8in}{!}{\includegraphics{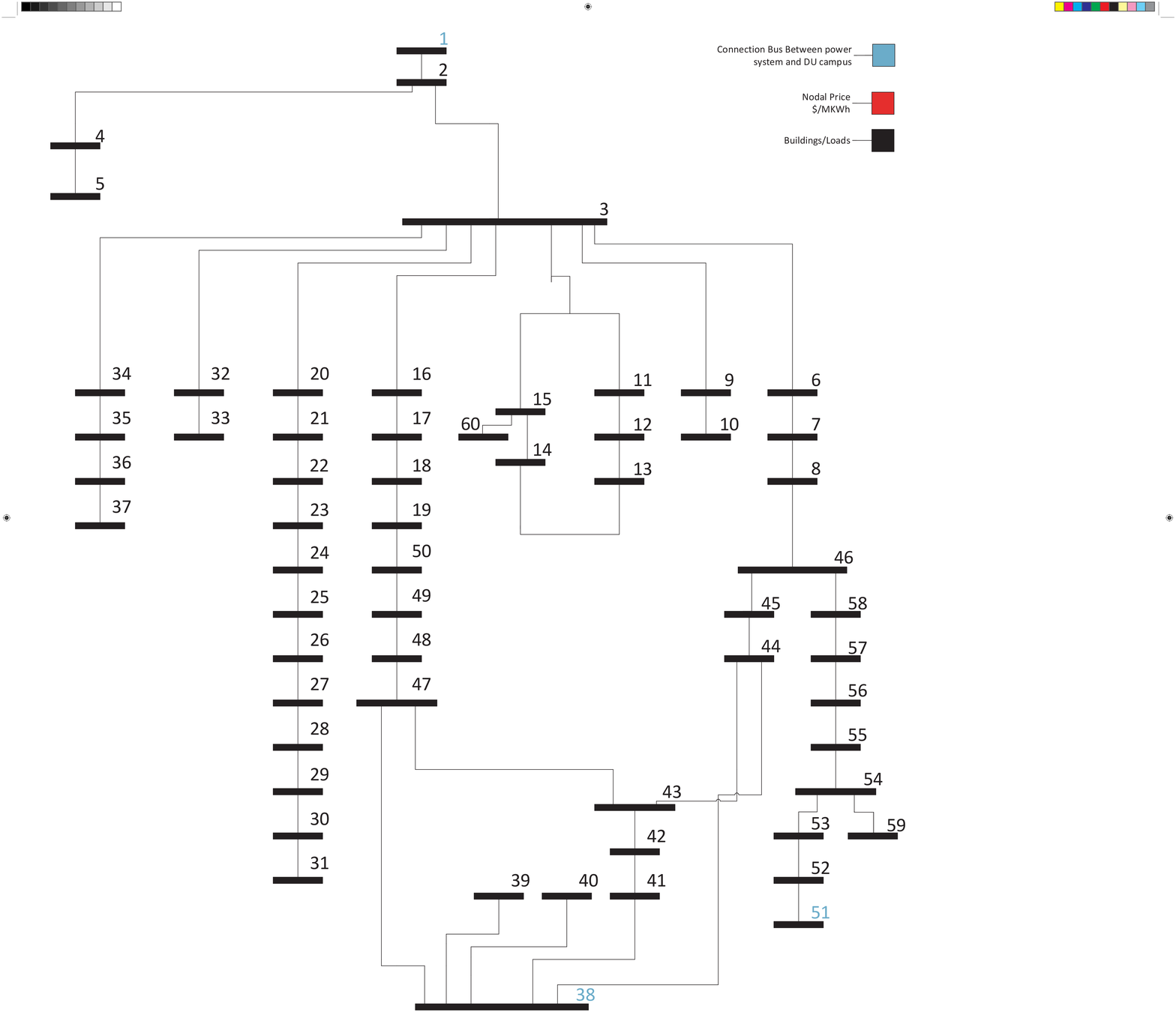}}}\\
		\subfigure[]{ \label{fig:simulink}
			\resizebox{2.8in}{!}{\includegraphics{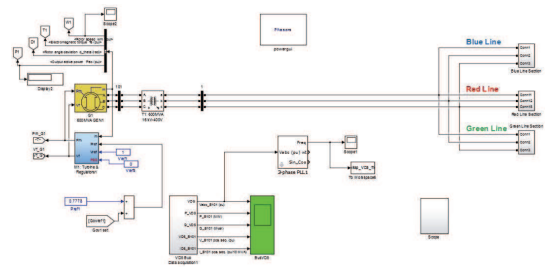}}}
		\caption{(a) The network topology of DU campus grid, (b) the corresponding test bed of DU campus grid.}
	\end{center}
\end{figure}

Fig.~\ref{fig:DULoopnetwork} demonstrates the network topology of the DU campus grid. In this network, the nodes stand for campus buildings and the lines represent the transmission lines between campus buildings. There are a number of switches on the campus, which are assumed to be closed in the numerical simulation. As shown in Fig.~\ref{fig:DULoopnetwork}, the total number of buses is 60. There are 57 buildings on campus and the other 3 buses are the set to represent the transmission lines. The power system is connected to the main utility grid at bus $1$, bus $38$ and bus $51$. The distribution renewable generators are assumed to be connected at bus $25$, bus $36$ and bus $42$. According to DU facility's Driscoll solar project, the University of Denver plans to spend \$$300\,k$ to assemble a $100\,kW$ photovoltaic generation. Correspondingly, it is assumed that three $40\,kW$ renewable generators are connected at each of the three buses. The rest of the buses are all load buses. The extracted power system topology can bring the feasibility of studying and simulating DLMP based on the DU campus grid system.

Fig.~\ref{fig:simulink} shows the corresponding simulation test bed we created in the Matlab Simulink environment. The campus power grid is divided into three sections according to the three different transmission lines in Fig.~\ref{fig:DULoopnetwork}. The DU electricity power profile is used to set the amount of the loads.

\section{The Locational Marginal Pricing at the Power Distribution Level}\label{chapter:4}
\subsection{Distribution Locational Marginal Pricing}
This paper implements an advanced approach of nodal pricing, which is an extension of LMP from transmission systems to distribution systems \cite{gu2017cha32432nce,gu2016knowledge}. Compared with transmission level, DLPM possesses its own distributional characteristics. In distribution power systems, the voltage is unified to all the buildings and the load may vary from time to time dramatically. Also in many distribution cases, the line congestion and flow limits are different from line to line. However, similarly to the LMP in transmission level, DLMP can be formulated with the following parts including marginal congestion cost (MCC), marginal loss cost (MLC) and marginal energy cost (MEC) \cite{jiang2017b123ig,jiang2012fa123ult,ding2016automa123tic}. This process is achieved in a distribution multi-source scheme, which will be explained in details later.

\begin{figure}[!t] 
	\begin{center}
		\includegraphics[scale=0.22]{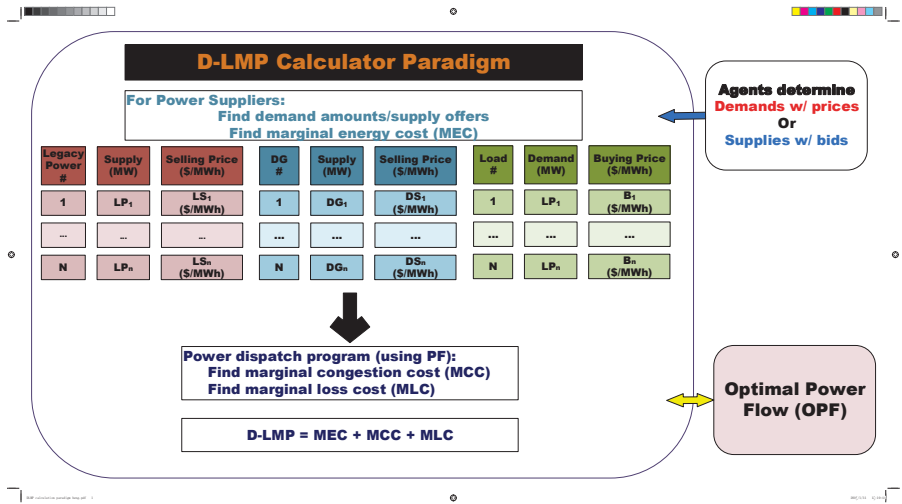}
		\caption{General block diagram of DLMP mechanism.}\label{fig:calculation paradigm}
	\end{center}
\end{figure}

The traditional scheme of distribution power systems is inflexible and will be outdated in the future. The conventional optimal power flow model corresponds to the minimization of the total cost of power production subject to power grid constraints. Nevertheless, in consideration of the forthcoming distribution power grids, with vast penetration of renewable energy and energy storage facilities, the electricity generation expenditure is not characterized clearly by far and it is difficult to find a well-recognized universal formulation. The DLMP takes advantages of social surplus as a substitution function in the OPF solver. In a competing power market, utility companies and consumers provide confidential offers and bids forecasting the expenditure and quantity that they are affordable to sell and buy electricity. As a result, the DLMP is enabled by such communication capability and the flow chart is shown in Fig.~\ref{fig:calculation paradigm}. The social surplus is depicted as the overall benefits of University of Denver subtracts the total costs of utility companies.

\begin{equation}\label{equ:maineq}
s=\sum\limits_{j=1}^{N}(c_j - p_j) \times q_{c_j} - \sum\limits_{i=1}^{M}(p_i - u_i) \times q_{u_i}
\end{equation}
where $s$ is the system social surplus that is gained from our DLMP calculation, $N$ is the total number of campus buildings and $j$ is the index of buildings; $M$ is the total number of electricity suppliers including renewable energy and $i$ is the index of those generations; $c_j$ stands the building bid price for each power generation and $u_i$ represents the offer price from each power generation; $p_j$ is the distribution locational marginal price at each building $j$, and $p_i$ stands for the distribution locational marginal price at supply bus $i$; $q_{c_j}$ is the power demand at building $j$; $q_{U_i}$ is the power supply from bus $i$.

\subsection{DCOPF Model for DLMP Calculation}\label{sec:DCmodel}
\subsubsection{Methodology}\label{sec:DC Methodology}
For a DC power system, the reactive power is not considered and the voltage magnitudes are set to be universal. In consideration of all the distribution power system characteristics, in the process of DLMP calculation, Generation Shift Factor (GSF) is used to reflect the time-varying line congestion corresponding to different line flow constraints \cite{meng2011distribution}. The implementation of real-time pricing mechanism is evaluated based on real DU campus power grid topology. It is assumed that in the future there are a number of renewable energy generations in the system. In terms of the DU campus power system, there are three legacy buses that can supply sufficient electricity from main utility grid to the campus. Optimal Power Flow (OPF) is the solver for our problem, and constraints are added into the traditional OPF problem to solve the DLMP problem. In this case, DCOPF is utilized to calculate DLMP. The constraints mainly consist of two parts: the balance between customer demands and supplier, and the constraints that includes $GSF_{k-i}$ to secure power transmission lines. As a result, the optimization problem can be modified as:
\begin{eqnarray}\label{equ:DCDLPM}
\underset{p_j, p_i}{\text{arg max}}&s=\sum\limits_{j=1}^{N}(c_j - p_j) \times q_{c_j} \nonumber\\
&- \sum\limits_{i=1}^{M}(p_i - u_i) \times q_{u_i}\\\label{equ:DCfunction}
\text{s.t.}&\sum\limits_{i=1}^{M}q_{u_i} - \sum\limits_{j=1}^{N}q_{c_j} = 0\\\label{equ:DCconstrain1}
&\sum\limits_{j=1}^{N} g_{k-i} \times (q_{u_i} - q_{c_j}) \leqslant f_k^{Max} \\\label{equ:DCconstrain2}
&q_{u_i}^{Min} \le q_{u_i} \le q_{u_i}^{Max}\\\nonumber \label{equ:DCconstrain3}
\end{eqnarray}
where $g_{k-i}$ is the generation shift factor from bus $i$ to line $k$, and $f_k^{Max}$ stands for the power flow limit at $k$th line.

The DC-DLMP problem can be divided into the following three components \cite{li2007dcopf}: 
\begin{eqnarray}
&p=MEC + MLC + MCC\\
&MEC = \lambda \\
&MLC = 0 \\
&MCC = \sum\limits_{i=1}^{N}g_{k-i} \times \mu_k
\end{eqnarray}
where $p$ is the distribution locational marginal price for each building; $MEC$ is the marginal energy cost, and $\lambda$ represents the Lagrangian multiplier of (\ref{equ:DCconstrain1}); $MLC$ is the marginal loss cost that equals to $0$ in DCOPF model; $MCC$ stands for the marginal congestion cost; $g_{k-i}$ is the generation shift factor, and $\mu_k$ is Lagrangian multiplier of (\ref{equ:DCconstrain1})

\subsubsection{Numerical Results on the DCOPF Based DLMP}\label{sec:test bed}
The DLMP based on DCOPF algorithm is simulated and evaluated in the DU 60-bus distribution system shown in Fig.~\ref{fig:DULoopnetwork}. The load configurations for each building are generated from real-world DU building data to create reasonable testing scenarios. The DU power system is assumed to consume all the available renewable energy to supply demands in order to reduce $CO_2$ emission and billing utility. Although the campus power grid utilizes all the renewable energy, it is not enough for balancing all the demands, and the campus power grid still needs supply from the legacy power grid, which means the campus will draw energy from buses $1$, $38$, $51$.

\begin{figure}[!htbp]
	\begin{center}
		\includegraphics[scale=0.29]{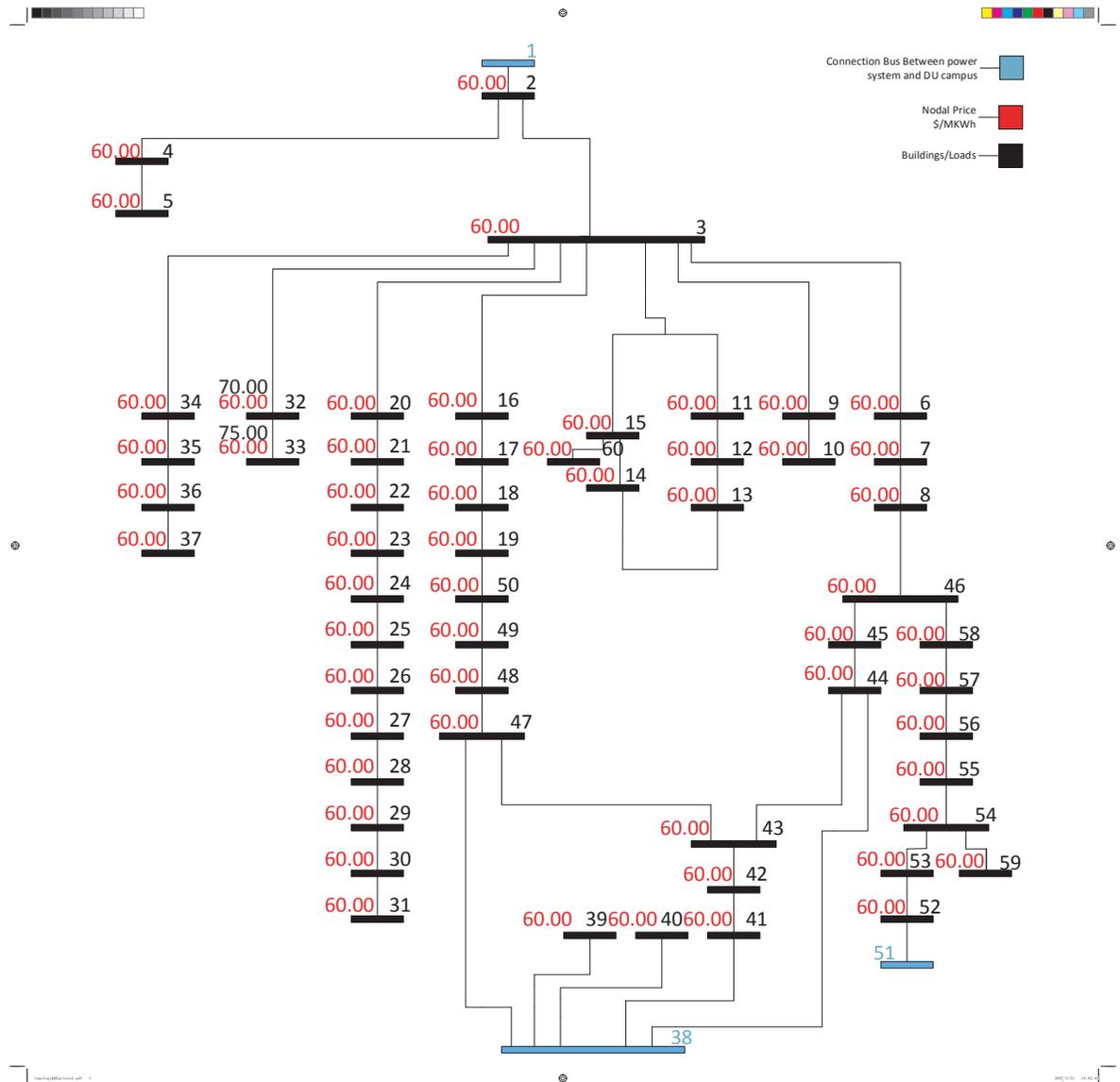}
		\caption{DC-DLMP calculation results.}\label{fig:DCDLPM}
	\end{center}
\end{figure}

In the numerical simulation, the overall load configurations of the DU campus power grid is $1879.98\,kW$ and $606.66\,kVar$ for active and reactive power, respectively. The DLMP calculation is developed in the MATLAB® environment based on the optimal power flow solver from MATPOWER 5.1 simulation package \cite{C4MATPOWER}. The simulation results are showed in Fig.~\ref{fig:DCDLPM}. The total power loss in the grid depends on load configurations. All the generation is dispatched based on DCOPF in order to maximize the social surplus on the DU campus.

As shown in Fig.~\ref{fig:DCDLPM}, the numbers in red show the prices for all the buses. The results shows that all the buildings in the DU campus power grid operate at the same price, $\$60.000$/MWh. Because in DCOPF model, the marginal loss ($MLC$) is assumed to equal zero and congestion loss ($MCC$) is also assumed to be zero as well, the overall power loss in DU campus distribution system is trivial. Congestion can influence the DLMP in the $DC$ model. In order to simulate the congestion condition, the ratings of the lines that are connected to bus 32 and bus 33 are decreased and the load of bus 33 is increased to 5 times of the normal load. The results corresponding to the modification is showed in black numbers in Fig.~\ref{fig:DCDLPM}. The prices for bus 32 and bus 33 are raised to $\$70.000$/MWh and $\$75.000$/MWh, respectively.

\subsection{ACOPF Based DLMP}
\subsubsection{Methodology}\label{sec:AC Methodology}
Compared with the DCOPF model, the solvers for loss and reactive power are added into ACOPF simulation, which makes the model include more distribution power system characteristics. In this paper, the DU campus grid is a medium-scale power system with real-world electricity profiles, which means the ACOPF model would not require a high computational load. The ACOPF based DLMP model is formulated as following.
\begin{eqnarray}\label{equ:ACDLPM}
\underset{p_j, p_i}{\text{arg max}}&s=\sum\limits_{j=1}^{N}(c_j - p_j) \times q_{c_j}\\\nonumber
&- \sum\limits_{i=1}^{M}(p_i - u_i) \times q_{u_i}\\\label{equ:ACconstrain1}
\text{s.t.} &\sum\limits_{i=1}^{M}q_{u_i} - \sum\limits_{j=1}^{N}q_{c_j} - L_{P}(V,\theta) = 0\\\label{equ:ACconstrain2}
&\sum\limits_{i=1}^{M}Q_{u_i} - \sum\limits_{j=1}^{N}Q_{c_j} - L_{Q}(V,\theta) = 0\\\label{equ:ACconstrain3}
&f_j(V,\theta) \leqslant f_j^{Max}\\\label{equ:ACconstrain4}
&q_{u_i}^{MIN} \le q_{u_i} \le q_{u_i}^{MAX}\\\label{equ:ACconstrain5}
&Q_{u_i}^{MIN} \le Q_{u_i} \le Q_{u_i}^{MAX}\\\label{equ:ACconstrain6}
&V_i^{MIN} \le V_i \le V_i^{MAX}\\\nonumber
\end{eqnarray}
where $V$ and $\theta$ are voltage magnitude and angle, respectively; $f_j$ stands for the power flow limit at $j$th line; $q_{u_i}$ is the active power output from each power source, while $Q_{u_i}$ is the reactive power output from the corresponding energy generation; $V_i$ stands for the voltage magnitude of the $i$th bus with power injection; and $L_{P}(V,\theta)$ and $L_{Q}(V,\theta)$ are the total active power loss and reactive power loss in the DU campus power gird, respectively.

\subsubsection{Numerical Results on the ACOPF Based DLMP}
\begin{figure}[!htb]
	\begin{center}
		\includegraphics[scale=0.29]{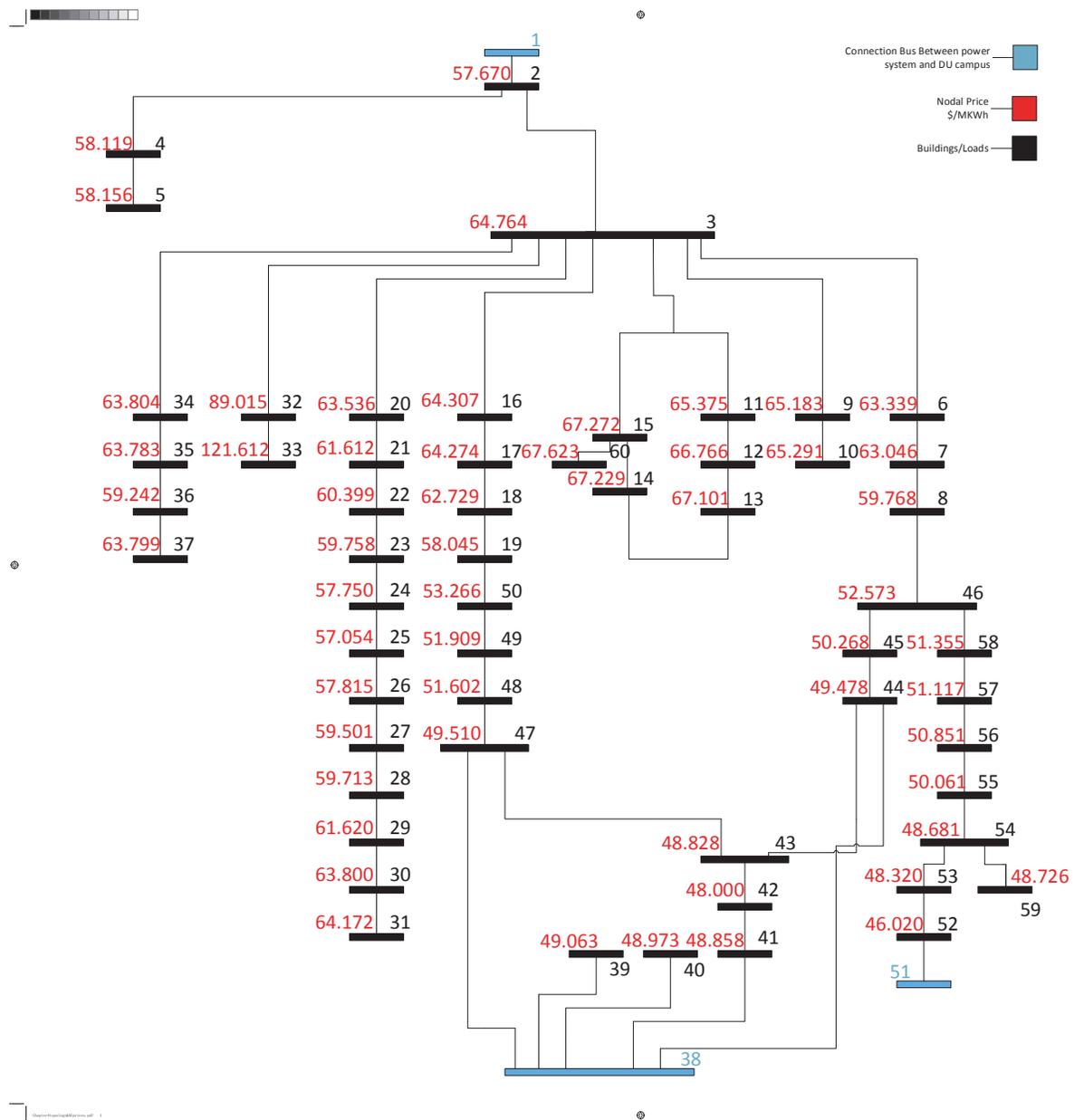}
		\caption{AC-DLMP calculation results.}\label{fig:ACDLPM}
	\end{center}
\end{figure}

In this ACOPF based DLMP model, the overall load configurations of DU campus power grid is set similarly to the DC model. The simulation results are depicted in Fig.~\ref{fig:ACDLPM}. The total active power losses is $78.856$ kW and reactive power losses is $20.06$ kVar. All the generation is dispatched based on the ACOPF results in order to maximize social surplus in DU campus.
The result shown in Fig.~\ref{fig:ACDLPM} indicates that in the ACOPF model all the buses have different prices. Building $33$ has the most expensive distribution locational marginal price, which is $\$121.612$/MWh. Building $52$ has the least expensive price, which is $\$46.020$/MWh. The average price of all the buildings is $\$59.379$/MWh.

Due to power loss, the generators need to provide extra energy to balance the load, therefore the additional marginal loss costs (MLC) are applied to each building. As a result, this makes the DLMPs more expensive than the corresponding DLMP values in DCOPF model. Also the overall social surplus calculated by ACOPF model is less than the total social surplus generated by DCOPF model.

\section{The Energy Consumption Model}
One of the major puzzles that all the power engineers and researchers are facing nowadays is that the access to real-time power profile can be hardly gained. Building managers and utility companies are very cautious when it comes to sharing the data to public. We can understand their trepidation because it would be changeling for them to share when they take security factors into consideration. Though it is very hard to get specific power consumption profile, there are actually several pretty good power simulation softwares, so we can utilize them and simulate the power profile according to our research and simulation purposes.

\subsection{The Building Consumption Analysis Tool}
To fit our research purpose, the building consumption analysis software should fulfill the following prerequisites: 1) Based on all kinds of buildings on campus, the consumption analysis tool can simulate various types of buildings, such as academic building, dormitory, arena \& fitness center and office buildings; 2) The consumption simulation software can provide up to hourly power profile because our research aims to optimize the hourly energy usage and power system operation and planning. 3) The building simulation tool should be capable of specify different architectures, number of story and the shape of the buildings.

\begin{figure}[!h]
	\begin{center}
		\includegraphics[width=0.7\textwidth]{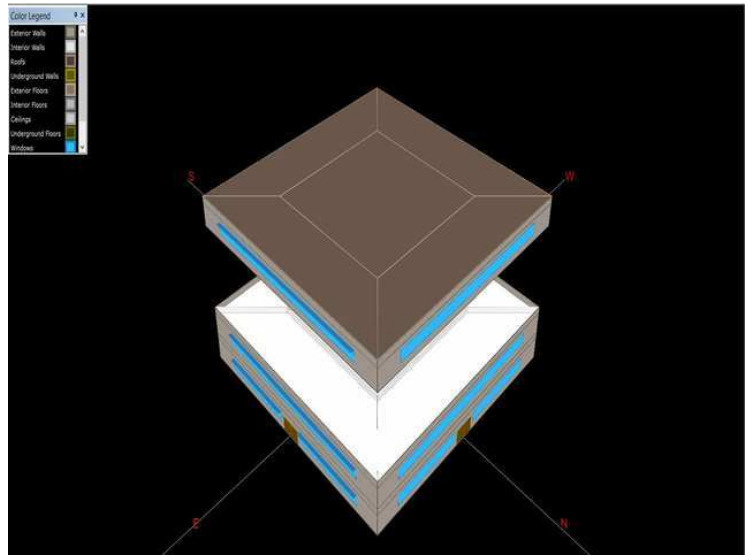}
		\caption{The building architecture design interface}
	\end{center}
\end{figure}

Among all the building consumption analysis tools, we choose eQUEST as our power profile provider. eQUEST is designed to provide whole building performance analysis to building professionals, i.e, owners, designers, operators, utility \& regulatory personnel, and educators \cite{2010equest}. eQUEST is a comprehensive and powerful tool to model building energy consumption, the analysis of building consumption is treated as system of systems, but the user interface and design process are designed to shorten and facilitate the process of preparing building models for simulation and research analysis. eQUEST is actually a window-based interface to the DOE simulation engine. The DOE-2.2 simulation engine is the most widely recognized, used, and trusted building simulation tool available today. And eQUEST have the ability that at the beginning of the establishment of the model, user has to specify the type of the building and the corresponding structure. If needed, researchers can also design the shape, material and amounts for the walls, windows and doors. 

\begin{figure}[!h]
	\begin{center}
		\includegraphics[width=0.7\textwidth]{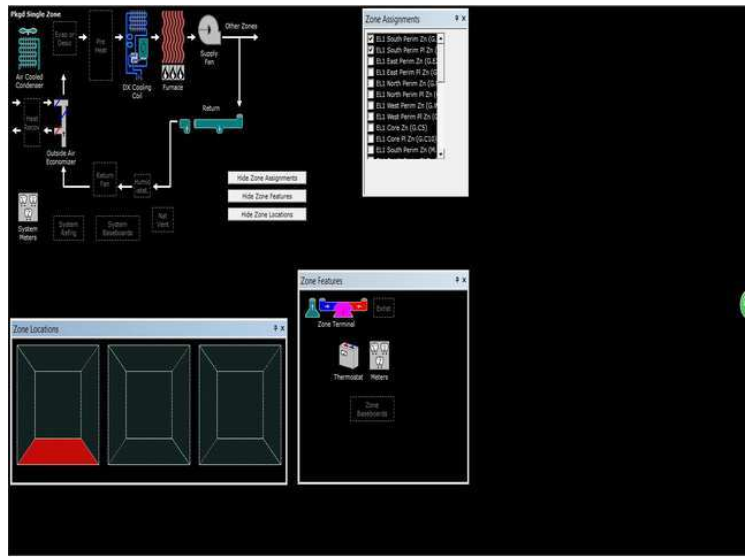}
		\caption{The HVAC system design interface}\label{eBuilding}
	\end{center}
\end{figure}

It should be noted that eQUSET can also estimate the HVAC system's size and model for the building that the user constructed in the software. Aside from that, Fig.~\ref{eBuilding} shows that the software can support complex system design and allows the researchers to design several independent HVAC system within one building. eQUEST is a simulation software that mainly aims to predict the energy consumption from HVAC system aspect, but it also provides the functionality to estimate the energy consumed by lights and plug-in load. This functionality makes the energy consumption predicted by eQUEST more applicable to real-world scenarios. Although it is impossible to get the same energy consumption as real-world power profile, we find that the output of eQUEST is good enough for our researches during the simulation process. And the output file contains detailed information about various types of data such as indoor temperature, outdoor temperature, date, lighting consumption and total end-use energy, etc.

\begin{table}[!h]
	\renewcommand{\arraystretch}{1.3}
	\caption{Partial Output Data}
	\label{tab:edata}
	\centering	
\begin{tabular}{|l|l|l|l|l|}
	\hline
	Lighting energy & Vent fan energy & Heating energy & Hot water energy & Total end-use energy \\ \hline
	23.6173         & 0               & 0              & 159479           & 159479               \\ \hline
	23.6173         & 0               & 0              & 160202           & 160202               \\ \hline
	23.6173         & 0               & 0              & 160732           & 160732               \\ \hline
	29.6397         & 0               & 0              & 161285           & 161285               \\ \hline
	104.204         & 397.022         & 1.74E+07       & 407303           & 1.78E+07             \\ \hline
	338.38          & 397.022         & 1.57E+07       & 1.82E+06         & 1.75E+07             \\ \hline
	413.313         & 397.022         & 1.50E+07       & 2.84E+06         & 1.78E+07             \\ \hline
	424.129         & 397.022         & 1.50E+07       & 2.29E+06         & 1.73E+07             \\ \hline
	425.112         & 397.022         & 1.48E+07       & 1.57E+06         & 1.64E+07             \\ \hline
	425.112         & 397.022         & 1.47E+07       & 1.35E+06         & 1.60E+07             \\ \hline
	425.112         & 397.022         & 1.41E+07       & 1.97E+06         & 1.61E+07             \\ \hline
	425.112         & 397.022         & 1.39E+07       & 3.19E+06         & 1.71E+07             \\ \hline
	425.112         & 397.022         & 1.37E+07       & 3.65E+06         & 1.73E+07             \\ \hline
	425.112         & 397.022         & 1.35E+07       & 2.37E+06         & 1.58E+07             \\ \hline
	425.112         & 397.022         & 1.33E+07       & 2.24E+06         & 1.55E+07             \\ \hline
	425.112         & 397.022         & 1.33E+07       & 3.55E+06         & 1.69E+07             \\ \hline
	425.112         & 397.022         & 1.35E+07       & 7.57E+06         & 2.11E+07             \\ \hline
	425.112         & 397.022         & 1.34E+07       & 1.41E+07         & 2.75E+07             \\ \hline
	425.112         & 397.022         & 1.33E+07       & 1.70E+07         & 3.03E+07             \\ \hline
	425.112         & 397.022         & 1.34E+07       & 1.54E+07         & 2.88E+07             \\ \hline
	414.786         & 397.022         & 1.37E+07       & 1.06E+07         & 2.43E+07             \\ \hline
\end{tabular}
\end{table} 
\chapter{Chapter 4}\label{Chap4}
\section{Introduction of Preliminary Study}
In this chapter, we provide a preliminary design to demonstrate the concept of social energy and parallel computing paradigm. The energy prediction models implemented in this chapter is a quadratic regression model and the objective function is solved by brute force searching algorithm. The goal of the part of research and work is to analysis the feasibility of the parallel computing paradigm introduced in Chapter~\ref{Chap2}, and lay a concrete foundation for further study.

\begin{figure}[!h]
	\centering
	\includegraphics[scale=0.34]{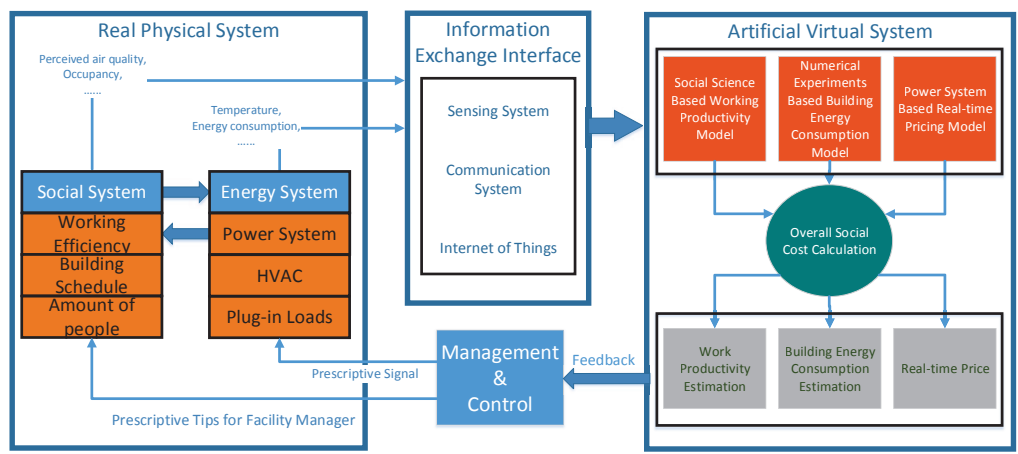}
	\caption{Blue print.}
	\label{fig:figx}
\end{figure}

The case study includes the elements of power system operation, smart building modeling, real-time pricing mechanism, and human behavior modeling. The technical scheme of the case study is demonstrated in Fig.\ref{fig:figx}, using the concept the parallel intelligence and control. The detailed technical description of the case study is introduced in this section. In previous literature, there are a lot of studies to prove the relationship between comfortness and work performance. Indoor temperature is a crucial factor of the indoor environment, which can affect human behavior in many ways such as perceived air quality, working performance and thermal comfort, etc. \cite{2006effect} The U.S federal government regulates CO$_2$ emission for universities, as a result, facility managers are required to meet the green house reduction regulation in order to avoid financial penalties by targeting on energy cutback. However, some of the strategies are inefficient and may cause reduction in comfortness. To some extent, to take residence who feel uncomfortable are less productive and students are less productive and need more time to accomplish their tasks, which may lead to more energy consumption and environmental degradation \cite{2008managing}. \cite{2004control,2004study} show that with \$2 saving per employer when indoor temperature is within comfortable range, the working efficiency will reduce 2\% per degree \degree C when the temperature is higher than 25\degree C.
\section{Preliminary Methodology}
This section proposes a smart building power consumption strategy by jointly considering the interactions between the campus power grid and the community artificial system. Work productivity is considered as one of the essential factor in the methodology, because work performance varies considerably under different indoor temperatures. Power consumption is related to indoor temperature settings and outside temperature. Based on \cite{2010equest}, a regression model is trained to simplify the calculation process for building power consumption. It should be noted that distributional marginal real-time pricing is implemented to reflect the changing of energy usage and utility expenses. The DLMP calculation is developed in MATLAB environment based on MATPOWER simulation toolbox \cite{2011matpower}.

\subsection{Human Work Performance Model}
Indoor temperature is one of the fundamental characteristics of the indoor environment. It can be controlled with a degree of accuracy depending on the building and its HVAC system. The indoor temperature affects thermal comfort, perceived air quality, sick building syndrome symptoms and performance at work \cite{2004control}. In this study, the work productivity $P$ is referred to the effects of temperature on performance at office work \cite{2006effect}. According to \cite{2002school}, the ideal temperature range for school is between 68\degree F and 74\degree F. Our target building is the fitness center in the University of Denver, therefore the temperature bracket is designed from 64\degree F and 79\degree F.
	
\begin{eqnarray}\label{equ:eff}
P=&0.1647524 \times ((T_{1}-32) \times 5/9)-\nonumber\\
&0.0058274 \times ((T_{1}-32) \times 5/9)^2+ \nonumber\\
&0.0000623 \times ((T_{1}-32) \times 5/9)^3-\nonumber\\
&0.4685328;\\
\text{s.t.}&64 \leqslant T_1 \leqslant79\\\nonumber
\end{eqnarray}
where $P$ is the work productivity, $T_{1}$ stands for the indoor temperature settings. (\ref{equ:eff}) will be used later to calculate the total building cost. Fig.~\ref{fig:eff} shows the relationship between indoor temperatures and the corresponding work efficiency.

\begin{figure}[!h]
\centering
\includegraphics[scale=1]{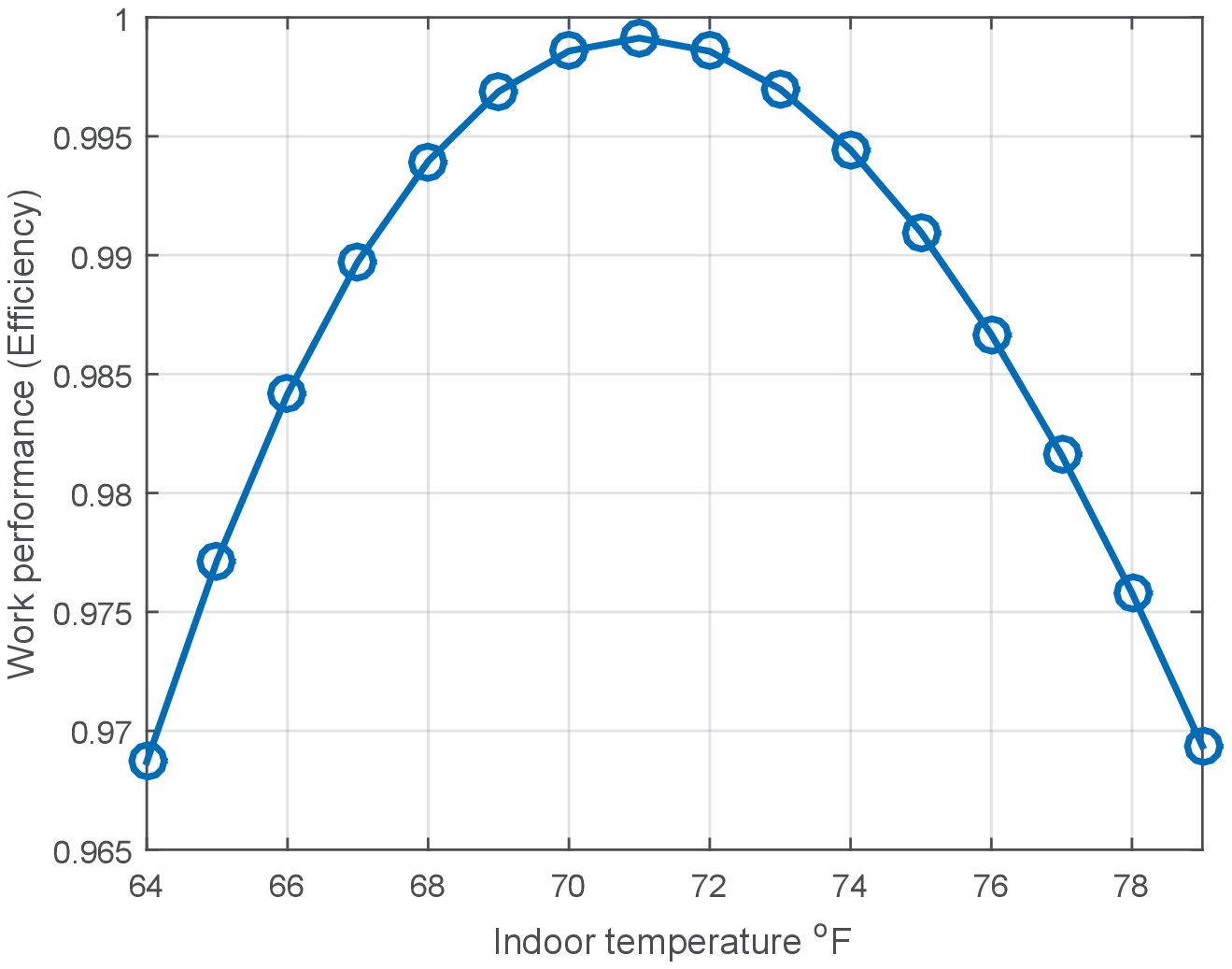}
\caption{Work performance w.r.t. indoor temperature.}
\label{fig:eff}
\end{figure}
	
\subsection{The Regression Model for Predicting Building Power Consumption}
We utilize \cite{2010equest} as the building simulation tool, which can provide comprehensive and detailed calculations about HVAC systems and simplistic assumptions for lighting and plug loads. The advanced hourly report can provide sufficient information for training the regression model for predicting power consumption. Daniel L. Ritchie center is the fitness center of the University of Denver (DU). More than 345,000 people visited this 41,000 m$^2$ building every academic year. In order to simulate the power consumption profile, the simulation model built in eQUEST is 41,000 m$^2$ as well. The type of the model is selected as fitness center. Table~\ref{tab:eQUEST1} shows part of the simulation results. It is noted that the simulation results generated much more information, results however, due to space limitation only the following is included in this paper. The fourth column shows the change of outside temperature.

\begin{table}[!h]
\renewcommand{\arraystretch}{1.3}
\caption{Part of the simulation results}
\label{tab:eQUEST1}
\centering
\resizebox{\columnwidth}{!}{
\begin{tabular}{|c|c|c|c|c|}
\hline
Month&Day&Hour&Temp (F)&Energy (BTU)\\
\hline
7 & 26 & 15 & 95 & $6.59 \times 10^6$\\
7 & 26 & 16 & 94 & $8.66 \times 10^6$\\
7 & 26 & 17 & 94 & $1.26 \times 10^7$\\
7 & 26 & 18 & 93 & $1.62 \times 10^7$\\
7 & 26 & 19 & 88 & $1.76 \times 10^7$\\
7 & 26 & 20 & 85 & $1.62 \times 10^7$\\
\hline
\end{tabular}
}
\end{table}

It is difficult to train a model that can fit the energy usage of all year round. Therefore, to ensure accuracy, the regression model in this paper focuses on the days of "cooling days". The designed indoor temperature varies from 64 \degree F to 79 \degree F, which is the same with temperature range in former section. Time, inside temperature and outside temperature are chosen from the simulation results to train the model. The model is shown in~\ref{equ:regress}.

\begin{eqnarray}\label{equ:regress}
E_n = &2.0443 \times t_1+1.8823 \times T_2\\\nonumber
&-1.6305 \times T_1+2.1181 \times 10^6\\
\text{s.t.}&10 \leqslant t_1 \leqslant 21\\
&64 \leqslant T_1 \leqslant 79\\
&50 \textless T_2 \textless 100\\\nonumber
\end{eqnarray}
where $E_n$ is the hourly consumed energy in british thermal unit (BTU), which is predicted by the pre-trained regression model; $t_1$ stands for the time; $T_2$ represents the outside temperature from the weather data; and $T_1$ is the indoor temperature.

\subsection{Distribution Locational Marginal Pricing for DU}
\begin{figure}[!h]
	\centering
	\includegraphics[width=0.9\textwidth]{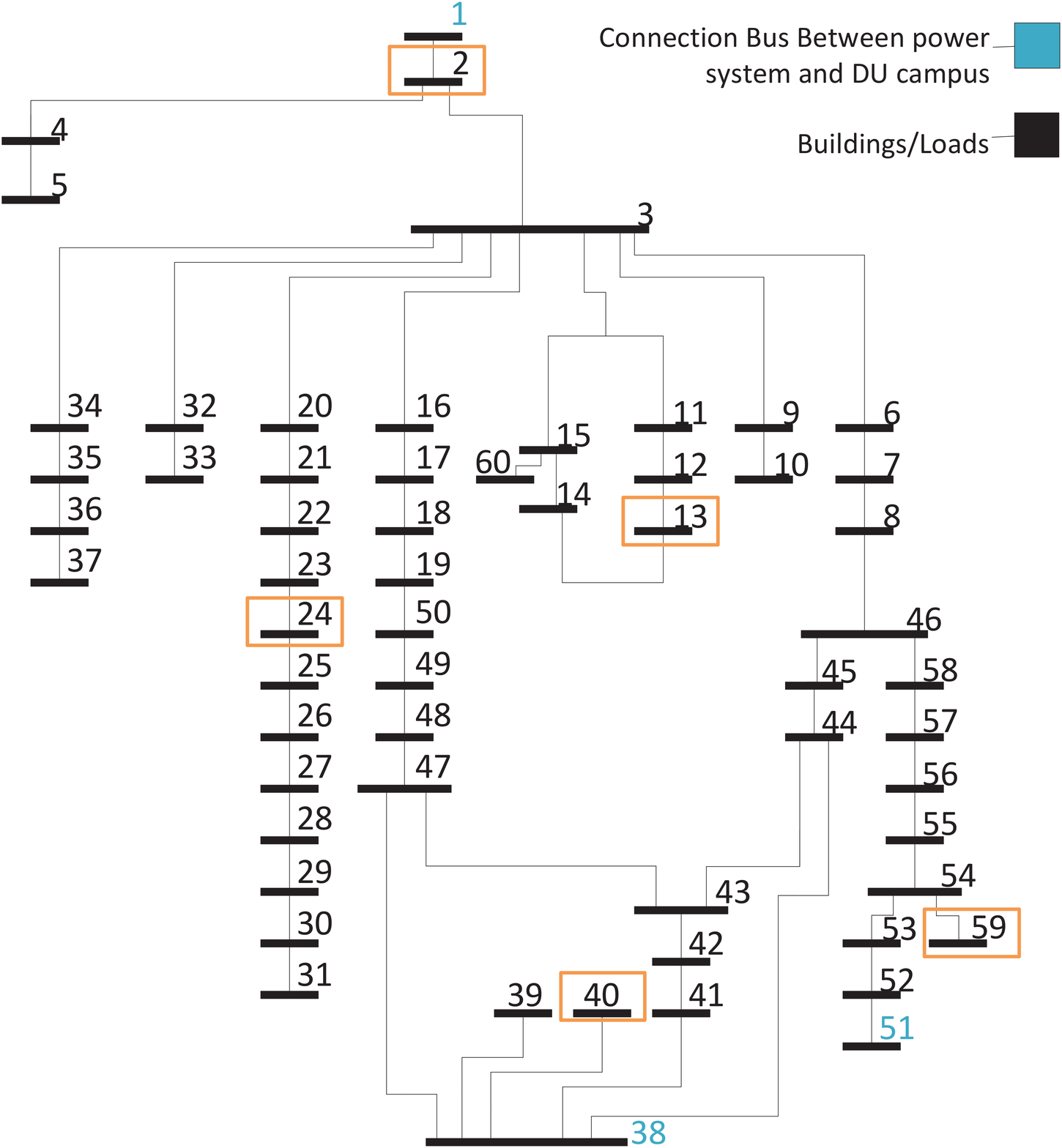}
	\caption{The network topology of DU campus grid.}
	\label{fig:topology}
\end{figure}

To indicate the influence of load changing on both financial aspect and campus power system side, distribution locational marginal pricing is introduced to associate the physical system to the artificial system. The detailed campus power system configurations is depicted in \cite{2016DU}. Fig.~\ref{fig:topology} shows the topology of DU campus power system. 

Using a simulator developed based on DU campus power system, the AC-OPF based DLMP is introduced to implement the real-time pricing mechanism.

\begin{eqnarray}\label{equ:ACDLPM}
\underset{p_j, p_i}{\text{arg max}}&s=\sum\limits_{j=1}^{N}(c_j - p_j) \times q_{c_j}\\\nonumber
&- \sum\limits_{i=1}^{M}(p_i - u_i) \times q_{u_i}\\\label{equ:ACconstrain1}
\text{s.t.} &\sum\limits_{i=1}^{M}q_{u_i} - \sum\limits_{j=1}^{N}q_{c_j} - L_{P}(V,\theta) = 0\\\label{equ:ACconstrain2}
&\sum\limits_{i=1}^{M}Q_{u_i} - \sum\limits_{j=1}^{N}Q_{c_j} - L_{Q}(V,\theta) = 0\\\label{equ:ACconstrain3}
&f_j(V,\theta) \leqslant f_j^{Max}\\\label{equ:ACconstrain4}
&q_{u_i}^{MIN} \leqslant q_{u_i} \leqslant q_{u_i}^{MAX}\\\label{equ:ACconstrain5}
&Q_{u_i}^{MIN} \leqslant Q_{u_i} \leqslant Q_{u_i}^{MAX}\\\label{equ:ACconstrain6}
&V_i^{MIN} \leqslant V_i \leqslant V_i^{MAX}\\\nonumber
\end{eqnarray}
where $s$ is the system social surplus that is gained from our DLMP calculation, $N$ is the total number of campus buildings and $j$ is the index of buildings; $M$ is the total number of electricity suppliers including renewable energy and $i$ is the index of those generations; $c_j$ stands the building bid price for each power generation and $u_i$ represents the offer price from each power generation; $p_j$ is the distribution locational marginal price at each building $j$, and $p_i$ stands for the distribution locational marginal price at supply bus $i$; $q_{c_j}$ is the power demand at building $j$; $q_{U_i}$ is the power supply from bus $i$; $V$ and $\theta$ are voltage magnitude and angle, respectively; $f_j$ stands for the power flow limit at $j$th line; $q_{u_i}$ is the active power output from each power source, while $Q_{u_i}$ is the reactive power output from the corresponding energy generation; $V_i$ stands for the voltage magnitude of the $i$th bus with power injection; and $L_{P}(V,\theta)$ and $L_{Q}(V,\theta)$ are the total active power loss and reactive power loss in the DU campus power gird, respectively.

All the other building will use time varying synthetic load to simulate the real-world campus power consumption condition.

\subsection{Overall Social Cost}
To address the overall social cost \cite{hao2016distribution,hao2017SMC}, a novel method is demonstrated in this section. Based on the aforementioned system configurations and social energy methodology, the formulation for the overall social cost comprises two major parts: the utility cost part, which is calculated by the end-use energy and the corresponding DLMPs; the cost of work productivity, which is determined by the cost of performance reduction and the amount of working personnels. (\ref{equ:final}) shows the formulation, and the goal of the equation is to search for the most economic combination between HVAC costs and work productivity.

\begin{eqnarray}\label{equ:final}
\underset{P_r, P, e_f}{\text{arg min}}&C = P_r \times E_n+E_f \times (1-P) \times O\\
\text{s.t.} & 0.96 \textless P \leqslant 1 \\
& 20 \textless P_r \leqslant 100 \\\nonumber
\end{eqnarray}
where $C$ is the overall cost; $P_r$ represents the DLMP at Ritchie Center; $E_n$ is the end-use energy; $E_f$ is the annual saving for each personnel when the working productivity is 1; $P$ stands for the actual working performance, which is determined by the indoor temperature settings; $O$ is the number of occupants in the building. Detailed results are discussed as following.

\subsection{Results and Discussion}
\subsubsection{Case 1: the Influence of Temperature settings}
\begin{figure}[!h]
	\centering
	\includegraphics[scale=0.9]{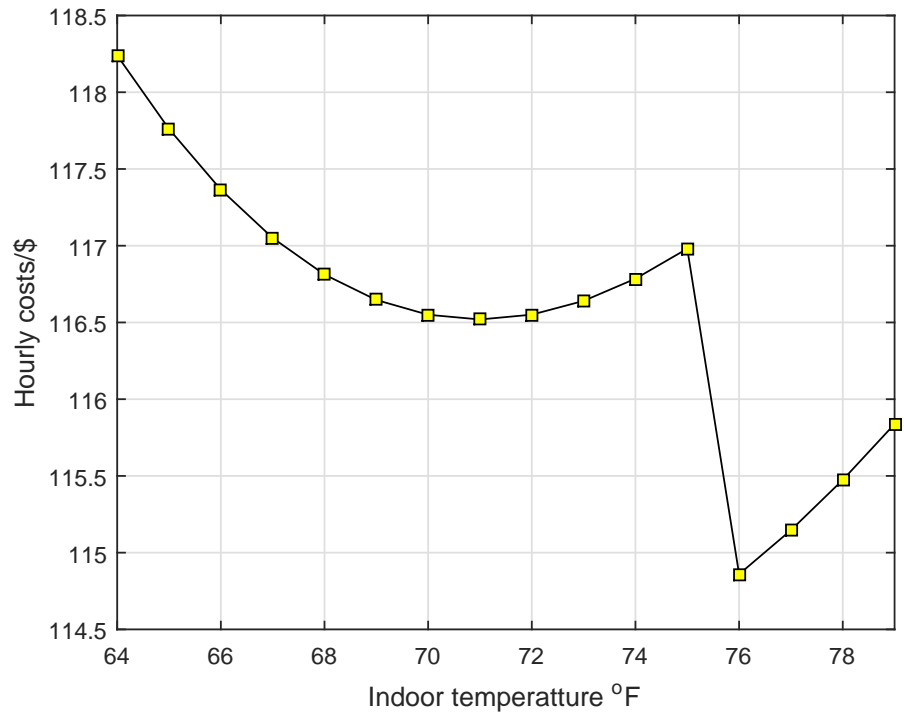}
	\caption{The influence of temperature settings.}
	\label{fig:case1}
\end{figure}

In this case, different temperature settings are tested when the time and amount of people are the same. According to the 2005-2006 academic year data, 345,000 people visited Ritchie Center, which means the average number is more than 900 people per day. It is assumed that the number of occupancy equals 500 in this case. 10 am is selected as the target time. The variable is the indoor temperature. As shown in Fig. ~\ref{equ:eff}, the best work performance occurs at 71\degree F, which equals to 0.9991. However, in Fig.~\ref{fig:case1}, it is suggested that the inside temperature should be adjusted to 76 \degree F. It should be pointed out that the resulting prices take considerations of both the utility cost and work performance cost. The hourly costs are \$114.86 and \$116.52, respectively. The hourly difference between those two settings are \$1.66, which can save more than \$6,000 per year. The most expensive temperature setting occurs at 64 \degree F, because the HVAC system needs to consume considerably energy to cool down the indoor temperature and the working efficiency is relatively low. Compared with the most expensive temperature setting, the suggested one can save up to \$12,337 per year, which is a significant saving. When the indoor temperature is set to 76 \degree F, there is a cost drop. The reason is that the needed cooling energy decreases and the working performance is relatively high. As shown, when the temperature is higher than 76 \degree F, the social cost is increasing due to the reduction of working productivity.
\subsubsection{Case 2: the Influence of Amount of People}
\begin{figure}[!h]
	\centering
	\includegraphics[scale=0.9]{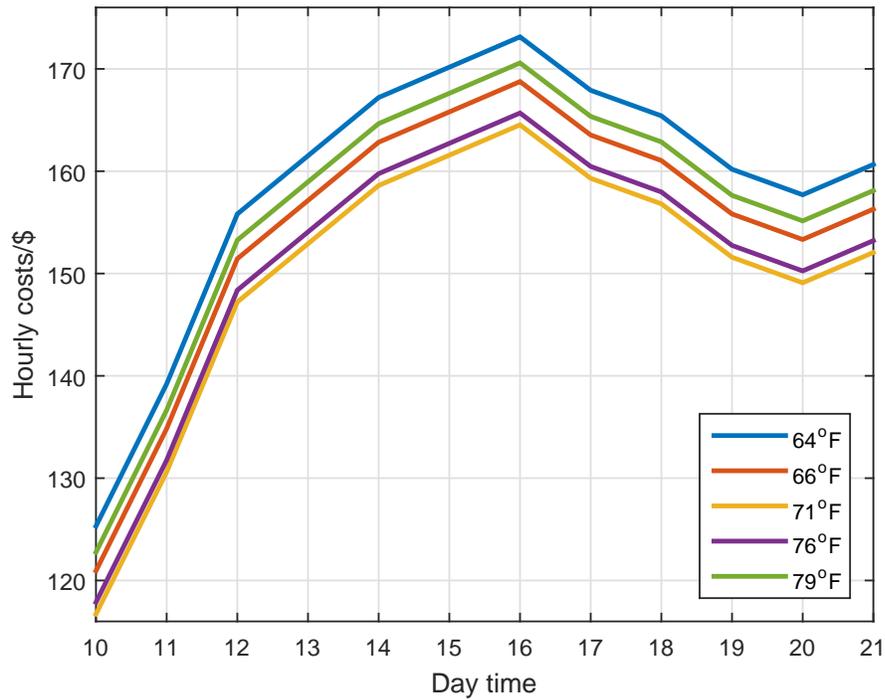}
	\caption{The influence of amount of people.}
	\label{fig:case2}
\end{figure}

During social events that happen in the Ritchie Center, such as sports games and commencements, the amount of people can increase dramatically. In this case, the number of people can probably be the major influence of the overall costs. As is demonstrated in (\ref{equ:final}), the total reduction of work efficient is influenced by the amount of people. In this scenario, it is assumed that 2,500 persons are inside the building. The various costs corresponding to different time and temperatures are demonstrated in Fig.~\ref{fig:case2}. Compared with Fig.~\ref{fig:case1}, the most economic temperature is 71 \degree F, which is the temperature for best work productivity. And early morning is much cheaper for DU to hold those kinds of events. In reality, lower temperature settings lead to more power consumption, but the setting can increase productivity.
\subsubsection{Case 3: the Comparison between Temperature and Amount of People}
\begin{figure}[!h]
	\centering
	\includegraphics[scale=0.9]{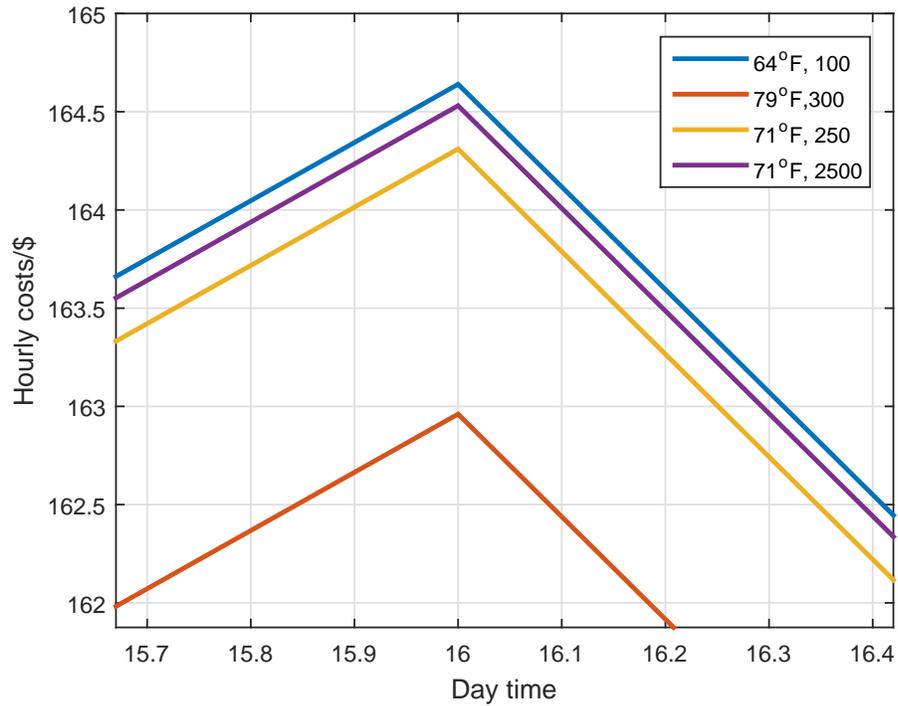}
	\caption{The comparison between temperature and amount of people.}
	\label{fig:case3}
\end{figure}

The comprehensive comparisons between different scenarios are conducted in this case study. Fig.~\ref{fig:case3} shows the total cost under various situations. The comparison mainly focuses on the most expensive time period in one day. Human beings generate thermal even when they are sleeping, therefore the amount of people in the building is related to the energy that needs to be consumed to cool down the room. However, in Fig.~\ref{fig:case3}, the real cost that serve 2500 people at 71\degree F is cheaper than providing a 64\degree F building for 100 personnels. And the overall costs for serving 300 people can lower than satisfying 250 people. The trade-off between work efficiency and temperature is measured and shown in this section. Detailed costs information is in Tab.~\ref{tab:cost} There is an old management adage: “What gets measured gets done.” The corollary is equally true: “What doesn’t get measured gets ignored.” Many facility managers only measure energy use. In this manuscript, the presented methodology can not only measure energy use but also measure the work performance, which can provide valuable maneuvers for facility managers to help them make decisions.

\begin{table}[!h]
	\renewcommand{\arraystretch}{1.3}
	\caption{Costs comparison}
	\label{tab:cost}
	\centering
	\resizebox{\columnwidth}{!}{
	\begin{tabular}{|c|c|c|c|}
		\hline
		Temp(F) & Personnels & Time & Costs(\$)\\
		\hline
		64 & 100 & 16 & 164.64 \\
		79 & 300 & 16 & 162.96 \\
		71 & 250 & 16 & 164.31 \\
		71 & 2500 & 16 & 164.53 \\
		\hline
	\end{tabular}
				}
\end{table}

\subsubsection{Case 4: One Day Simulation}
\begin{figure}[!h]
	\centering
	\includegraphics[scale=0.9]{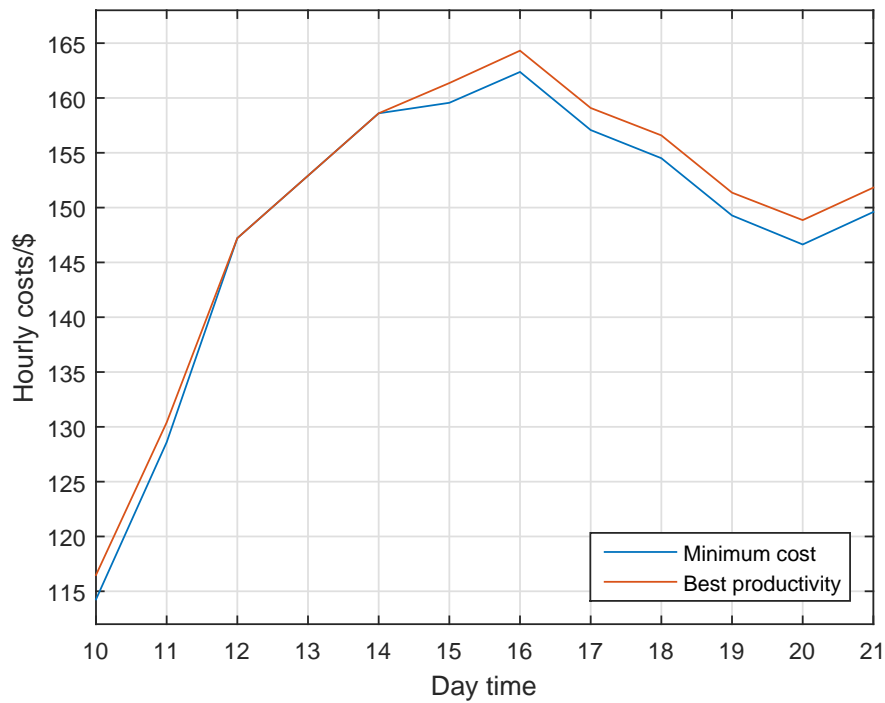}
	\caption{The one day hourly cost comparison.}
	\label{fig:case4}
\end{figure}

As aforementioned, Ritchie Center is built as a multi-functional building. DU holds many kinds of games and commencements during whole year. In this section, it is assumed that a event will be held in Ritchie center from noon to 2 pm and the amount of people is up to 2,500. From the results of previous case studies, the total social cost is pretty high under certain temperature circumstances. 71\degree F is not only the best temperature setting for working productivity but also a practical temperature setting in real life. Therefore, the hourly cost of 71\degree F is 
set as the referral in this section. As shown in Fig. \ref{fig:case4}, except for the time period from noon to 2 pm, the minimum cost (the blue line) is lower than the referral curve (the red line). Compared with the referral, the proposed methodology can help to save \$18.40 per day or \$6716 per year. The detailed hourly information is demonstrated in Tab.~\ref{tab:cost1}. The second column is the amount of people inside the building. The third column is the minimum cost calculate by the proposed method. The fourth column is the referring cost.

\begin{table}[!h]
	\renewcommand{\arraystretch}{1.3}
	\caption{Costs comparison}
	\label{tab:cost1}
	\centering
	\resizebox{\columnwidth}{!}{
	\begin{tabular}{|c|c|c|c|}
		\hline
		Time & Occupancy & Minimum (\$) & 71\degree F (\$)\\
		\hline
		\hline
		10 & 100 & 114.26 & 116.48 \\
		\hline
		11 & 400 & 128.59 & 130.40 \\
		\hline
		12 & 2,500 & 147.21 & 147.21 \\
		\hline
		13 & 2,500 & 152.91 & 152.91 \\
		\hline
		14 & 2,500 & 158.60 & 158.60 \\
		\hline
		15 & 400 & 159.56 & 161.36 \\
		\hline
		16 & 300 & 162.37 & 164.32 \\
		\hline
		17 & 250 & 157.10 & 159.10 \\
		\hline
		18 & 200 & 154.51 & 156.59 \\
		\hline
		19 & 200 & 149.28 & 151.36 \\
		\hline
		20 & 100 & 146.64 & 148.86 \\
		\hline
		21 & 100 & 149.60 & 151.82 \\
		\hline
	\end{tabular}
}
\end{table}
	
\section{Section Conclusion}
In this section, a preliminary methodology to calculate the costs of social energy is demonstrated. This innovative formulation is based on the paradigm of parallel computing. The methodology can be the cornerstone for the following researches that studies the combination costs of electricity and work efficiency. 
\chapter{Chapter 5}\label{Chap5}
\section{The Problem and Motivation}
The methodology introduced in Chapter \ref{Chap4} mainly aims to demonstrate the novel idea we proposed in this manuscript. During the later research process and the development of modern artificial intelligence and neural network technologies, we realize that the methodology in Chapter \ref{Chap4} possesses several shortcomings as listed in the followings:
\begin{itemize}
	\item Limitation of energy consumption model: the energy prediction model in Chapter ~\ref{Chap4} is trained by linear regression method which limits the capability of the algorithm. The limitation of linear regression model is that it can only predict a certain time period in one day, for example, 10AM to 9PM in summer. There is an urgent demand to implement an advanced method to train a sophisticated model that can predict the power profile for entire year.
	\item Brute force searching method: the preliminary results shows in Chapter ~\ref{Chap4} is calculated by exhaustive search. The restriction of the proposed searching methodology is that it cannot be achieved in any large scale power system. When the number control strategy and buses increases, the calculation complexity and time will escalate exponentially.
	\item The methodology needs to be tested and examined in more complicated and comprehensive cases and scenarios.
\end{itemize}

At the same time, through the literature of Chapter \ref{Chap2}, Chapter \ref{Chap3} and Chapter \ref{Chap4}, we can know that the demand for design an algorithm that can fully fulfill the following characteristics is vital.

\begin{itemize}
	\item The algorithm have the ability to yearly assist the power system planning and operation and be able to adapt to the continuously changing environment and customer energy usage.
	\item The method is capable of stimulating building managers to participate because future smart grid will integrate many features such as renewable energy, real-time pricing and distributed generator. Even in the same campus, but each building may have various energy usage demand according to their diverse control goals.
	\item The approach should offer the building manager the best control strategy corresponding to the miscellaneous building's types and schedules.
	\item Not like Brute force searching method, the algorithm should be capable of handling the calculation for large scale power systems, a distributed or multi-agent methodology may be able to cope the problem.
	\item The new methodology needs to be proved and examined in more complicated and comprehensive cases and scenarios.
\end{itemize}

As society progresses and technology develops, the power system becomes more and more complicated, and humans’ requirements and expectations upon the system have been leveled up step by step. The phases of power system improvements start with the successful delivery of energy, to safe energy delivery, to expansion for larger coverage and capacity, to advancements in stability and resilience, to improvement of energy efficiency, and to enhancement of social welfare. The developments of renewable energy resources in most of universities or societies are relatively slow, because the solar panels and wind turbines are expensive Most buyers cannot afford using renewable energies. Meanwhile, the integration of renewable energy needs the power grid to have more modularity and and adaptability which may reduce the system robustness and uncertainty in demands and generations. On the other hand, the integration of renewable energy further fluctuates the real-time pricing (RTP) in future smart grid. Therefore, for energy end users like universities, price is one of the crucial factors for cost saving and load reduction \cite{SGresidential}. In \cite{SGEVcharging}, authors present a novel method for energy exchange between electric vehicle (EV) and power grid based on the Stackelberg game. However, for academic and commercial buildings, the impact of EV is negligible in terms of the amount of load consumed by EVs. \cite{SGpricingmech} concentrates on the design and the implementation of game theory in RTP and load stability for future power grid. The paper introduces social welfare for maximizing the supplier and consumer profits. Yet, the study on the influence of RTP was not included. Some researchers conducted experiments about the relationship between RTP and user activities \cite{SGjhgPrc1,SGjhgPrc2}. The price mechanisms used in those articles, common flat rate and quadratic cost function for example, will not be suitable for future smart grid. Energy scheduling and demand management will benefit smart gird in many aspects including large scale load shift, congestion mitigation, reduction of power system transit stability issues.

Social cost, which includes the electricity consumption costs and human working efficiency costs, is introduced as an advanced concept to address the importance of both the energy consumption and human experiences in power system management. The optimization of social cost is designated as the objective function to arrange and manage the HVAC system scheduling. Inspired by the methodology in \cite{gamemethod}, we transform the aforementioned problem into a game-theoretic problem. The proposed approach can solve a finite n-person non co-operative game that the players are the building managers in the same local smart grid, the strategies are the HVAC indoor temperature settings, and the payoffs are the social cost according to various personnel inside those buildings and the indoor working productivity. It should be noted that distributional locational marginal pricing (DLMP) is introduced to strengthen and reflect the relationship among the player payoffs, other player's action and power system situation. To illustrate the proposed methodology, we embedded the approach into an interactive real-artificial parallel computing technique. For implementing our methodology and the artificial-real  management/computing system, human efficiency modeling, numerical experiments based smart building modeling, distribution level real-time pricing and social response to the pricing signals are studied. The details of aforementioned techniques will be depicted in the following sections.

The rest of the chapter is organized as the following: Sec.\ref{sec:models} describes the computing and communication systems and the corresponding modules; Sec.\ref{sec:game} narrates the formulation of the social game; Sec.\ref{sec:results} shows the simulation results and the related analytical description. Sec.\ref{sec:conclusion} concludes the chapter.

\section{Establishment of Computing System}
\label{sec:models}
In this section, we provide analytical descriptions of the proposed computing and communication mechanism, the working productivity model, the energy consumption model and the power system DLMP model. We assume that the buildings is equipped with smart meters with the capability of two-way communication. According to each building schedules the smart management system can forecast the amount of people inside the buildings. 

\subsection{Real-artificial Computing and Management Paradigm}
\begin{figure}[!h]
	\centering
	\includegraphics[width=1.00\linewidth]{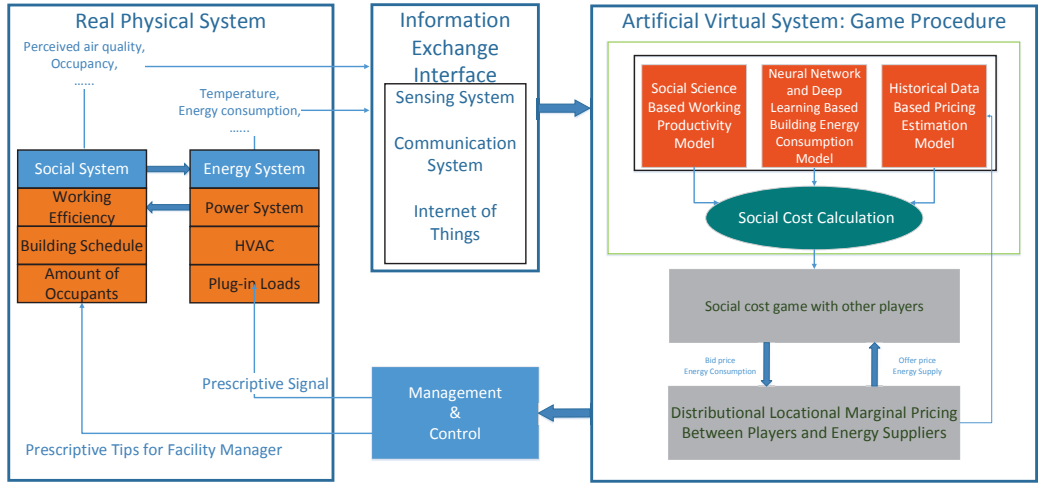}
	\caption{Technical scheme of the DU social energy system case study.}
	\label{fig:TechFrame}
\end{figure}

Our proposed methodology is to minimize building energy consumption and suggest the building manager the best HVAC settings to ensure that the indoor environment is within a comfort zone. To fulfill the function ality of this methodology, we introduce the real-artificial computing and management paradigm to be the computational and communicational framework based on the concepts in \cite{parallelsystem}. Fig.~\ref{fig:TechFrame} shows the real-artificial computing and management paradigm.

Our optimization methodology requires the information interaction between the physical systems such as HVAC system and the digital systems like smart meter. The real-artificial computing and management paradigm consists of two major parts: the real physical system and the artificial virtual system. The physical system collects data such as the amount of people inside a building, the building schedule for next time period and the current indoor temperature setting. Using the information collected by physical system, the virtual system will be able to calculate the payoff matrix and game with other players in the same local smart grid. Through optimization, the virtue system will provide feed-back to the building manager to help set the optimal HVAC temperature for the next time period.

\subsection{Human Work Performance Model}
\label{sec:human}
Indoor temperature is one of the fundamental characteristics of the indoor environment. It can be controlled with a degree of accuracy depending on the building and its HVAC system. The indoor temperature affects thermal comfort, perceived air quality, sick building syndrome symptoms and performance at work \cite{2004control}. The work productivity $P$ is referred to effect of temperature on performance at office work \cite{2006effect}, and the human work performance model can be expressed as
\begin{eqnarray}\label{equ:eff}
\xi &=& g(T_{in})\\
&=& 0.1647524 \cdot (\frac{5}{9} \cdot (T_{in}-32))-\nonumber\\
&&  0.0058274 \cdot (\frac{5}{9} \cdot (T_{in}-32))^2+ \nonumber\\
&&  0.0000623 \cdot (\frac{5}{9} \cdot (T_{in}-32))^3-0.4685328 \nonumber
\end{eqnarray}
where $\xi$ is the work productivity, $T_{in}$ is the indoor temperature settings which satisfies $T_{l} \leqslant T_{in} \leqslant T_{u}$ and $T_{l}$ is the temperature lower limit, $T_{u}$ is the temperature upper limit. And Fig.~\ref{fig:eff} shows the relationship between the indoor temperatures and the work efficiency. It should be pointed out that, although according to \cite{2002school}, the ideal temperature range for university buildings is between 68\degree F and 74\degree F, the temperature bracket in our study is designed between 64\degree F and 79\degree F.
\begin{figure}[!h]
	\centering
	\includegraphics[width=1.0\linewidth]{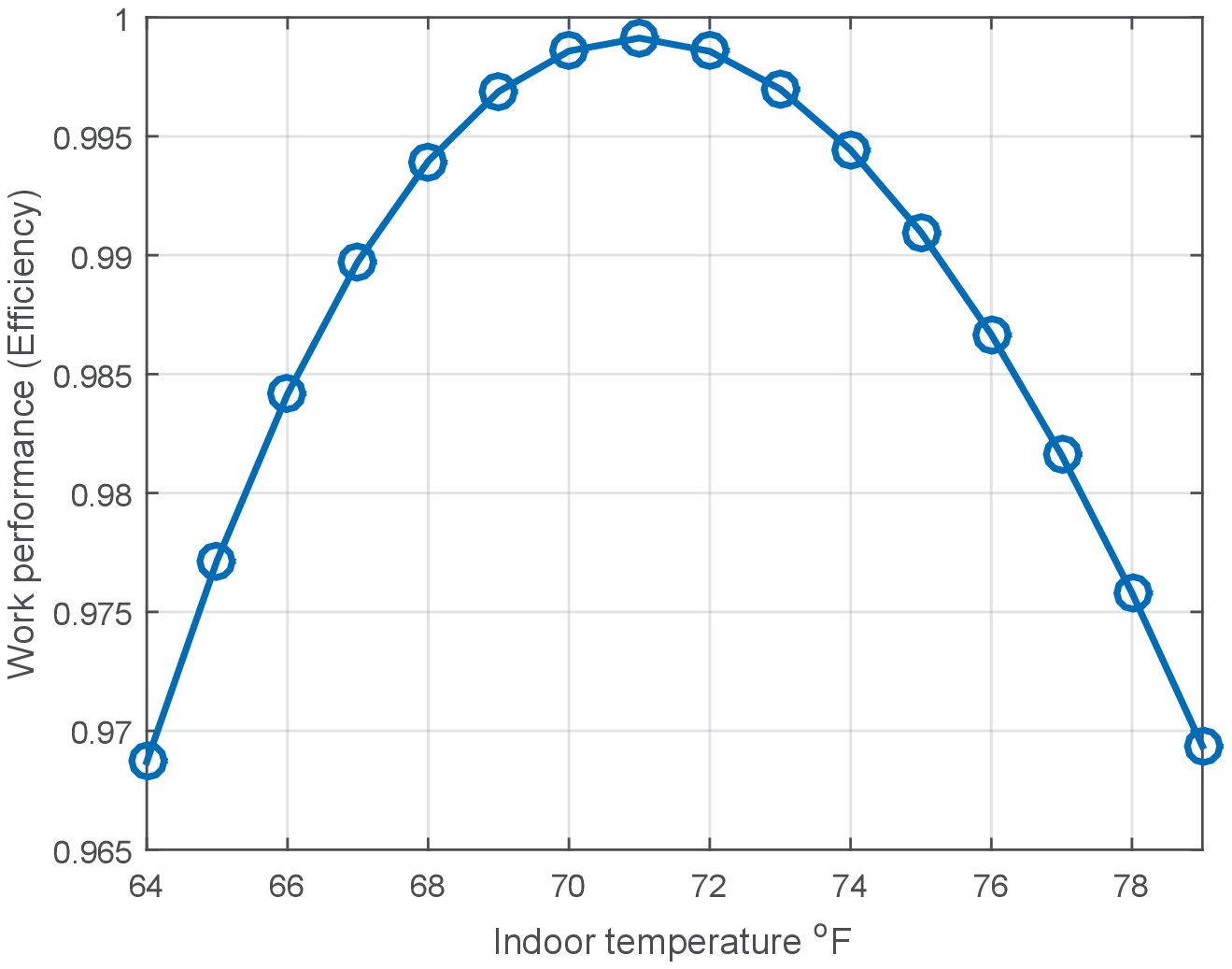}
	\caption{Work efficiency as a function of indoor temperature.}
	\label{fig:eff}
\end{figure}
Let $\xi_{k,t}$ and $x_{k,t}$ denotes the work productivity and indoor temperature setting in building $k$ at time $t$, respectively, (\ref{equ:eff}) can be rewritten as
\begin{eqnarray}\label{equ:efffunc}
\xi_{k,t} = g(x_{k,t}).
\end{eqnarray}
In addition, we denote $\boldsymbol{\xi}_t=[\xi_{1,t} \; \xi_{2,t} \dots \xi_{n,t}]^\intercal$ and $\mathbf{x}_t=[x_{1,t} \; x_{2,t} \dots x_{n,t}]^\intercal$, $\mathbf{x}_t$ is the control variable in this case study.

\subsection{The Neural Network Based Energy Consumption Profile Models}
\label{sec:consumption}
We utilize eQUEST \cite{2010equest} as the building simulation tool, which provides comprehensive and detail calculations about HVAC systems and simplistic assumptions for lighting and plug-in loads according to the size and type of various buildings. The hourly report from the simulation software can provide sufficient information for training the neural network models to predict power consumption. Compared with previous chapters, the capability and accuracy of the building energy consumption model is improved to the next level. Table~\ref{tab:eQUEST} shows partial simulation results of the hourly energy consumption on July $1st$ at the Ritchie center. 

\begin{table}[!h]
	\renewcommand{\arraystretch}{1.1}
	\caption{Partial simulation results of Ritchie center on July $1st$ from $15:00$ to $20:00$}
	\label{tab:eQUEST}
	\centering
	\begin{tabular}{|c|c|c|c|}
		\hline
		Hour&T$_{in}$ (\degree F)&T$_{out}$ (\degree F)&Energy (BTU)\\
		\hline
		15 &64& 94 & $1.08\times10^7$\\
		16 &64& 92 & $1.31\times10^7$\\
		17 &64& 92 & $1.70\times10^7$\\
		18 &64& 93 & $2.03\times10^7$\\
		19 &64& 89 & $2.22\times10^7$\\
		20 &64& 85 & $2.19\times10^7$\\
		\hline
	\end{tabular}
\end{table}

Simulated data in one year under different indoor temperature settings should be generated for each building in the same smart grid. 
For each building, a two-layer feed-forward network with sigmoid hidden neurons and linear output neurons is trained using the Levenberg-Marquardt backpropagation algorithm. Hour, indoor temperature setting and outdoor temperature are the inputs, and the energy consumption is the output of the neural network model. In our study, 75\%, 15\% and 15\% of the sample data are randomly selected as the training, validation and testing data, respectively. 
The average R-value is 0.92, which demonstrates that the neural network models for predicting energy consumption are acceptable. Thus, energy consumptions of building $k$ at time $t$, which is denoted as $e_{k,t}$ can be expressed as functions of indoor temperature setting $x_{k,t}$, time $t$ and outdoor temperature $T_{out,t}$
\begin{eqnarray}\label{equ:consump}
&e_{k,t} = h_k (x_{k,t}, t, T_{out,t})\\
&\mathbf{e}_t = H (\mathbf{x}_t, t, T_{out,t}) \nonumber
\end{eqnarray}
where $\mathbf{e}_t=[e_{1,t} \; e_{2,t} \cdots e_{57,t}]^\intercal$. 

\subsection{Distribution Locational Marginal Pricing for DU Campus Grid}
\label{sec:price}
To indicate the influence of load variation on both financial aspect and campus power system side, the distribution locational marginal pricing \cite{DLMPmethod} is introduced to associate the physical system with the artificial system.
\begin{eqnarray}\label{equ:ACDLPM}
\underset{p^b_j, p^g_i}{\text{arg max}}&s=\sum\limits_{j=1}^{N}(c_j - p^b_j) \cdot q_{c_j}\\\nonumber
&- \sum\limits_{i=1}^{M}(p^g_i - u_i) \cdot q_{u_i}\\\label{equ:ACconstrain1}
\text{s.t.} &\sum\limits_{i=1}^{M}q_{u_i} - \sum\limits_{j=1}^{N}q_{c_j} - L_{P}(V,\theta) = 0\\\label{equ:ACconstrain2}
&\sum\limits_{i=1}^{M}Q_{u_i} - \sum\limits_{j=1}^{N}Q_{c_j} - L_{Q}(V,\theta) = 0\\\label{equ:ACconstrain3}
&f_j(V,\theta) \leqslant f_j^{Max}\\\label{equ:ACconstrain4}
&q_{u_i}^{MIN} \leqslant q_{u_i} \leqslant q_{u_i}^{MAX}\\\label{equ:ACconstrain5}
&Q_{u_i}^{MIN} \leqslant Q_{u_i} \leqslant Q_{u_i}^{MAX}\\\label{equ:ACconstrain6}
&V_i^{MIN} \leqslant V_i \leqslant V_i^{MAX}\\\nonumber
\end{eqnarray}
where $s$ denotes system social surplus that is obtained from our DLMP calculation, $N$ is the number of buildings in smart grid and $j$ is the building index; $M$ is the total number of electricity suppliers and $i$ is the generator index; $c_j$ stands for the building bid price for each power generation and $u_i$ represents the offer price from each power generation; $p^b_j$ is the distribution locational marginal price at each building $j$, and $p^g_i$ stands for the distribution locational marginal price at supply bus $i$; $q_{c_j}$ is the power demand at building $j$; $q_{u_i}$ is the power supply from bus $i$; $V$ and $\theta$ are voltage magnitude and angle, respectively; $f_j$ stands for the power flow at $j$th line, which is limited by  $f_j^{max}$ A; $q_{u_i}$ is the active power output from each power source and the maximum capacity $q_{u_i}^{MAX}$ MW, while $Q_{u_i}$ is the reactive power output from the corresponding energy generation and the maximum capacity $Q_{u_i}^{MAX}$ MVar; $V_i$ stands for the voltage magnitude of the $i$th bus with power injection, in this case study $V_i^{MIN}$ pu and $V_i^{MAX}$ pu; and $L_{P}(V,\theta)$ and $L_{Q}(V,\theta)$ are the total active power loss and reactive power loss in the smart gird, respectively.

For the convenience of calculations, we express the DLMP of the target buildings $\mathbf{p}_t = [p_{1,t} \; p_{2,t} \dots p_{n,t}]^\intercal$ as a highly nonlinear and complex function $\Gamma(\cdot)$ of energy consumption $\mathbf{e}_t$, and thus as a function of the control variable $\mathbf{x}_t$ as
\begin{eqnarray}\label{equ:dlmpfunc}
\mathbf{p}_t = \Gamma (\mathbf{e}_t) = \Gamma({H(\mathbf{x}_t, t, T_{out_t})}).
\end{eqnarray}

\subsection{Overall Social Cost}
To address the overall social cost, a novel method is demonstrated in this section. Based on the aforementioned real-artificial computing and management paradigm and the system configuration, the formulation for the overall social cost comprises two major parts: the utility cost, calculated by the end-use energy and the corresponding DLMPs; the cost of work productivity, determined by the cost of performance reduction and the amount of occupants. (\ref{equ:scost}) defines the formulation for calculating the overall social costs $\psi_t$ at time $t$ of the target buildings in smart grid.
\begin{eqnarray}\label{equ:scost}
\psi_t & = & \sum_{k = 1}^{n} [p_{k,t} \cdot e_{k,t} + w \cdot \alpha (1-\xi_{k,t}) \cdot o_{k,t}]\\ \nonumber
& = & \mathbf{p}_t \cdot \mathbf{e}_t + w \cdot \alpha (\mathbf{1}-\boldsymbol{\xi}_t) \cdot \mathbf{o}_t\\ \nonumber
& = & \Gamma [H(\mathbf{x}_t,t,T_{out,t})] \cdot H(\mathbf{x}_t,t,T_{out,t}) \\ \nonumber
&   & + w \cdot \alpha[1-g (\mathbf{x}_t)] \cdot \mathbf{o}_t\\
& = & \Psi (\mathbf{x}_t, t, T_{out,t}, \mathbf{o}_t) 
\end{eqnarray}
where $\psi_t$ is the overall cost at time $t$, $w$ is the weight for the efficiency component which is set to be $0.1$, $\alpha$ is the hourly saving for each personnel when the working productivity is 1， and $\mathbf{o}_t = [o_{1,t} \; o_{2,t} \cdots o_{n,t}]^\intercal$ where $o_{k,t}$ is the number of occupants in building $k$ at time $t$. 

It should be noted that, the outdoor temperature $T_{out,t}$ and the number of occupants $\mathbf{o}_t$ is obtained through historical data and is directly related to the time $t$ in a day. For this reason, at a certain time $t$, $T_{out,t}$ and $\mathbf{o}_t$ are known. The overall social cost is a function of the indoor temperature settings $\mathbf{x}_t$. The goal is to find, the best indoor temperature settings $\hat{\mathbf{x}}_t$ at time $t$ that generates the most economic combination between HVAC costs and work productivity. Therefore, the problem can be formulated as
\begin{eqnarray}\label{equ:mincost}
\hat{\mathbf{x}}_t &=& \underset{\mathbf{x}_t}{\text{arg min}} \; \Psi (\mathbf{x}_t) \\ \nonumber
& \text{s.t.} & T_{l} \leqslant x_{k,t} \leqslant T_{u}
\end{eqnarray}

\section{Social Cost Game}
\label{sec:game}
Based on the models and the controlling and management mechanism depicted in Sec.\ref{sec:models}, optimizing the setting of HVAC system can be formulated. Buildings consume up to $45\%$ of the global energy, in it, HVAC system constitutes the largest user of energy. Therefore the optimization of building HVAC system energy usage is crucial not only to our environment but also to the occupants. Our proposed methodology solves a finite N person game, since the game is designated to maximize the social cost within a certain smart grid while means that the players are the buildings managers that each of the individual manager aims to minimize the energy cost and maintain the indoor temperature in reasonable zone in a smart grid. According to the temperature control bracket and accuracy of HVAC system, the number of strategies for each player is finite. The payoffs are affected by utility price which is presented in Sec.\ref{sec:price}, the HVAC consumption whose energy usage is modeled in Sec.\ref{sec:consumption} and working efficiency of building occupants is related to the indoor temperature, and analyzed in Sec.\ref{sec:human}.

We devise the social cost optimization problem into a $N$ person, finite non co-operative game which can be expressed in following:
\begin{equation}\label{eq:game1}
\psi(x_{i,t}) = (\mathbf{N},\{\mathbf{S}^{i,t}\}_{i\in{N}},\{\mathbf{\delta}^{i,t}\}_{i\in{N}})
\end{equation}
where $\mathbf{N}$ is the player set at time $t$, $\mathbf{S}^{i}$ is the $i_{th}$ players' finite pure strategy set at time $t$ and $\mathbf{\delta}^{i}$ is the payoff set corresponding to various pure strategies at time $t$. If the game has an optimal solution, there is at least one Nash equilibrium, which means that one player could not get a better payoff than the optimal strategy at NE if the game reaches a Nash equilibrium and the other players are playing according to their Nash equilibrium strategies.
\begin{equation}
\delta^{i,t}(\alpha^*) \geqslant \delta^{i,t}(\alpha^{*-i},\alpha^i), \forall i \in N, \forall \alpha^i \in \Sigma^i
\end{equation}
where $*$ means the strategy at NE, $-i$ denotes the players other than $i_{th}$ player, and $\Sigma^i$ is the mixed strategy space for player $i$ 

However, the DLMP mechanism narrated in Sec.\ref{sec:price} is a double auction pricing paradigm and the nodal prices are calculated by alternating current optimal power flow(AC-OPF). If players keep communicating with energy sellers during the period of the games, the whole process would be incredibly long making the feasible algorithm into an infeasible one. Therefore, we implement the price estimation tool from \cite{SGprc} for the managers to estimate their nodal prices and payoffs. An effective way to predict is likely by referring to the prices at the same time from yesterday, the day before yesterday, and the same day last week. 
Therefore, the predicted price for an upcoming hour $t$ on a day $d$ is obtained as followed:
\begin{equation}\label{eq:conspric}
\hat{p}^t[d] = k_1p^t[d-1]+k_2p^h[d-2]+k_7p^t[d-7],\forall h\in H.
\end{equation}
Here, $p^t[d-1]$, $p^t[d-2]$, and $p^t[d-7]$ denote the previous values of price $p^t$ on yesterday, the day before yesterday, and the same day last week, respectively. We choose All-Days-Same Coefficients that means the parameters $k_1$,$k_2$,$k_7$ remain the same for every day \cite{SGprc}, and we set $k_1=0.8765$, $k_2=0$, $k_7=0.1025$. After implementing those coefficients, when compared with DLMPs, the price prediction error resulting from (\ref{eq:conspric}) is $16\%$ on average. After the managers reach NE, their energy usage scheduling information will be sent to the DLMP module in artificial system, then the virtual system will calculate the actually DLMPs and feed back the nodal prices to the managers through the physical system. The results from DLMPs will be substituted for $[d-1]$,$[d-2]$, and $[d-7]$ sequentially until the difference of prices between the manger estimation and the DLMPs is within the predetermined price threshold $\epsilon$.  While the pricing is updating, each manager will also update their strategy and game with others.

\begin{algorithm}
	\label{alg:alg1}
	\begin{algorithmic}[1]
		\State Random initialization
		\State Based on (\ref{eq:conspric}) calculate social cost and payoff matrix.
		\State \textbf{Repeat}.
		\State Each player take turns and update payoff-matrix. 
		\State Calculate best response (BR) strategies.
		\State According to BR, implement DLMP to calculate price.
		\State Each player broadcast the updated price and payoff matrix.
		\State \textbf{End}
		\State Update schedule for next time interval.
	\end{algorithmic}
\end{algorithm}
By assuming $\psi^i$ is the optimal HVAC system setting for player $i$, then we have an nonlinear optimization problem for the each player in smart grid. 
\begin{eqnarray}\nonumber
&(\psi^{i,t})\\
&\text{min}&\gamma^{i,t}-\delta^{i,t}(\alpha)\\
&\text{s.t.}&\delta^{i,t}(\alpha^{-i,t},s^{i,t}_j)-\gamma^{i,t} \leqslant 0 \forall j = 1,\dots,m^i\\
&&\sum_{j=1}^{m^i}\alpha^{i,t}_j = 1\\
&&\alpha^{i,t}_j \geqslant 0 \quad \forall j = 1,\dots,m^i 
\end{eqnarray}
where $\gamma^{i,t}$ is assumed as the optimal social cost corresponding to the best indoor temperature setting, $(\alpha^{-i,t},s^{i,t}_j)$ denotes the player $i$'s strategies set include strategy $s^{i,t}_j$ while the others' strategies are expressed as $\alpha^{-{i,t}}$ at time $t$. According to \cite{gamemethod}, after applying KKT condition, we can obtain that a Nash equilibrium of game (\ref{eq:game1}) can be transformed into a problem of equalities and inequalities.
\begin{lemma}\label{lemma}
	A necessary and sufficient condition for game $\psi$ to have a Nash equilibrium strategy set $\alpha$ is
	\begin{eqnarray}\label{eq:lemma1}
	&\gamma^{i,t}-\delta^{i,t}(\alpha) = 0 \quad \forall i \in N\\\label{eq:lemma2}
	&\delta^{i,t}(\alpha^{-i,t},s^{i,t}_j)-\gamma^{i,t} \leqslant 0, \\\nonumber
	&\forall j = 1,\dots,m^i,\forall i \in N\\\label{eq:lemma3}
	&\sum_{j=1}^{m^i}\alpha^{i,t}_j = 1,\forall i \in N\\\label{eq:lemma4}
	&\alpha^{i,t}_j \geqslant 0 \quad \forall j = 1,\dots,m^i, \forall i \in N
	\end{eqnarray}	
\end{lemma}
Form (\ref{eq:lemma1}), we can obtain that for every player in the same smart grid their best response strategy is at Nash equilibrium. (\ref{eq:lemma2},\ref{eq:lemma3},\ref{eq:lemma4}) are the equality and inequality constraints for optimization and (\ref{eq:lemma2}) means that no mix strategy combination would result in better social cost than best response. Therefore, we can obtain that the optimal solution of nonlinear HVAC scheduling problem is the strategy at Nash equilibrium $\alpha$.  
\begin{theorem}
	A necessary and sufficient condition for $\alpha^*$ to a Nash  equilibrium of game $\Psi$ is that it is an optimal solution of the following minimization problem 
	\begin{eqnarray}\nonumber
	&(\Psi)\\\nonumber
	&\text{min}&\sum_{i \in N}\gamma^{i,t}-\delta^{i,t}(\alpha)\\\nonumber
	&\text{s.t.}&\delta^{i,t}(\alpha^{-i,t},s^{i,t}_j)-\gamma^{i,t} \leqslant 0,\\\nonumber
	&& \forall j = 1,\dots,m^i, \forall i \in N\\\nonumber
	&&\sum_{j=1}^{m^i}\alpha^{i,t}_j = 1, \quad \forall i \in N \\\nonumber
	&&\alpha^{i,t}_j \geqslant 0 \quad \forall j = 1,\dots,m^i, \forall i \in N 
	\end{eqnarray}
\end{theorem}
The optimal value of our proposed social cost game should be 0. The value of $\gamma^{i,t}$ at the optimal point gives the expected payoff of the player $i$ at time $t$.
\subsection{Simulation Results}\label{sec:results}
\subsubsection{Simulation Testbed}
Like the previous chapters, the University of Denver campus power grid is used as the simulation testbed in this chapter. There are 60 buses on campus, $57$ buses are load bus and the other $3$ buses are power sources. Assume that all the buildings on campus have been equipped with smart meters which have two-way communication capability using a campus local network and the proposed social game algorithm has been deployed. The campus power consumption varies between $2$ MW and $11$ MW. For those three power sources, we set the maximum power supply for active power and reactive power to be $12.7$ MW and $11$ MVar, respectively. The bus voltage are limited by $V_i^{MIN} = 0.90$ pu and $V_i^{MAX}=1.10$, $i\in N = 60$. A detailed campus power system configurations is depicted in \cite{2016DU}. Fig.~\ref{fig:topology} shows the topology of DU campus power system. 
\begin{figure}[!h] \hspace{-20pt}
	\centering
	\includegraphics[width=0.9\textwidth]{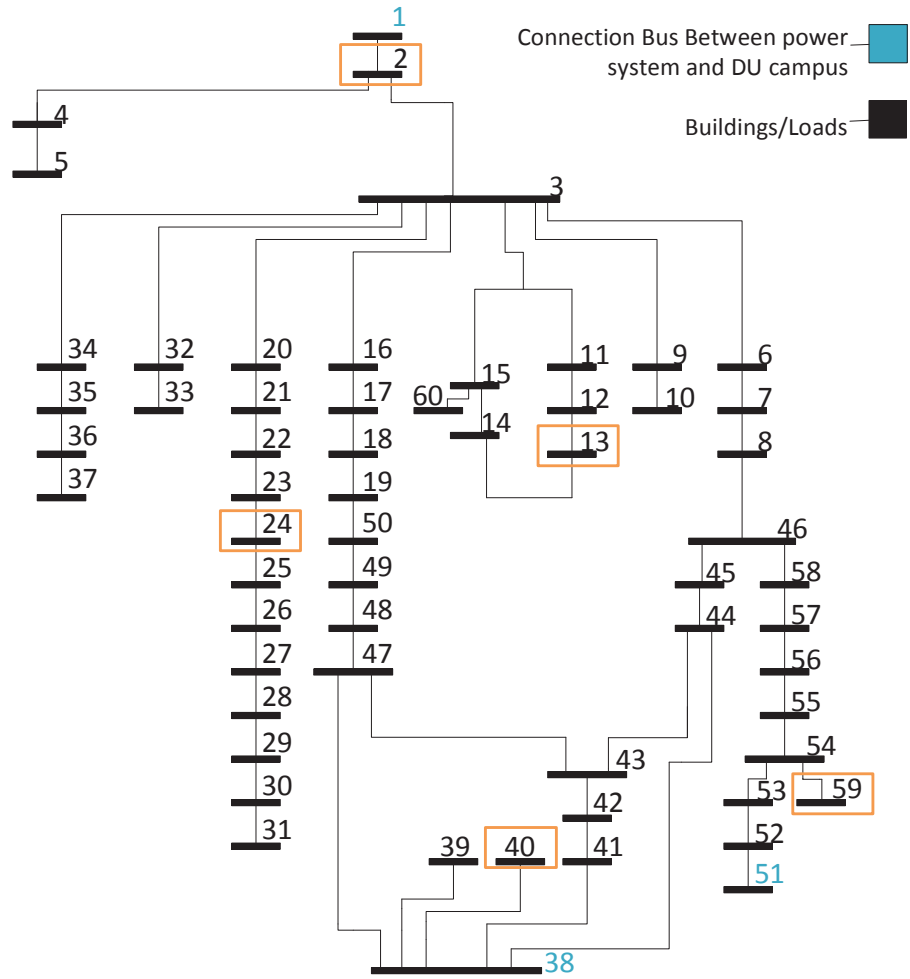}
	\caption{The network topology of DU campus grid.}
	\label{fig:topology5}
\end{figure}
\subsubsection{Results}\label{sec:gamer}
\begin{figure}[h]
	\centering
	\includegraphics*[width=0.7\linewidth]{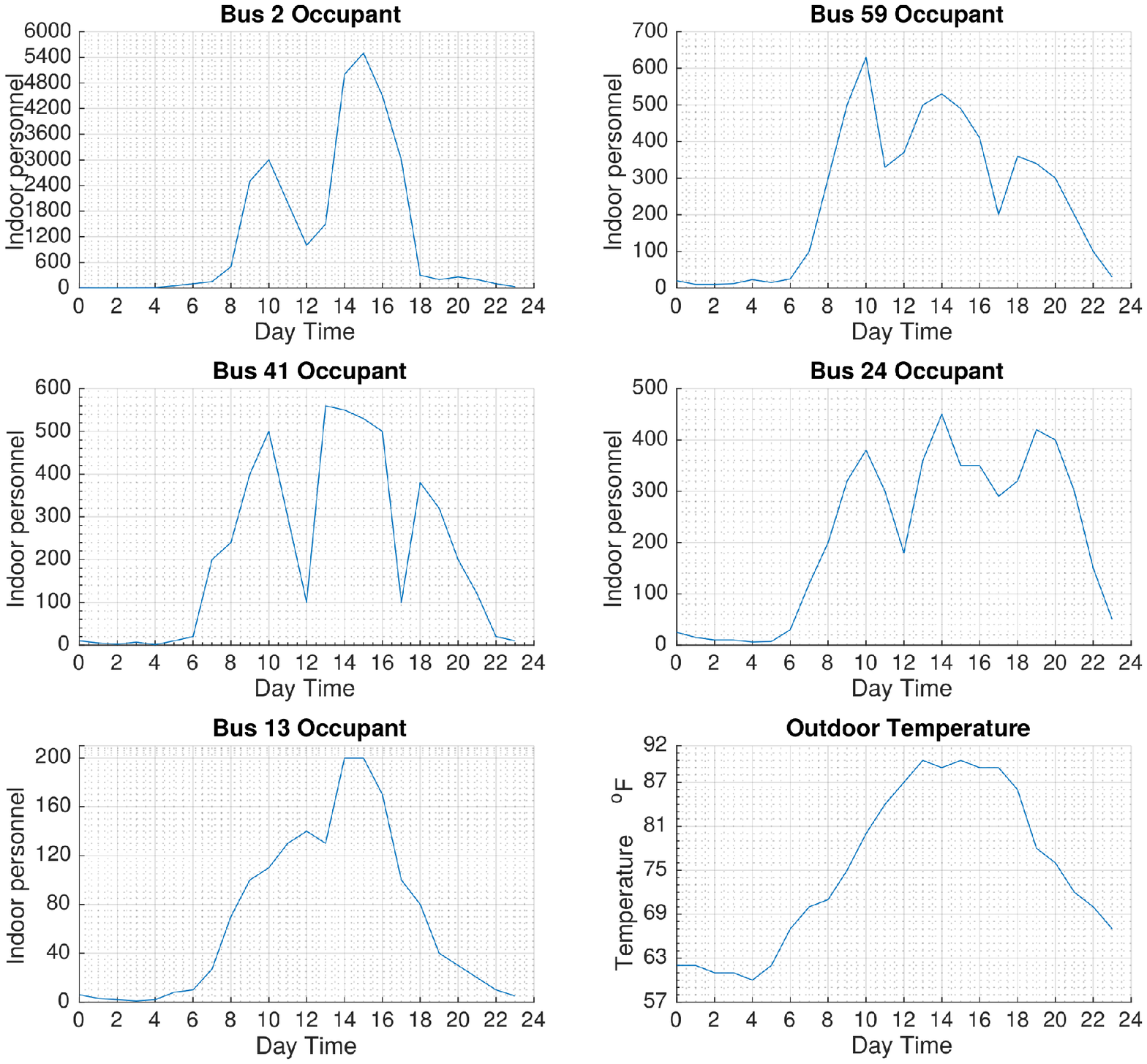}
	\caption{The amount of people inside buildings}
	\label{fig:amoutpeople}
\end{figure}

In this chapter, bus $2$, bus $59$, bus $41$, bus $24$, and bus $13$ are selected as the social cost game players and are highlighted in Fig.\ref{fig:topology}. Those buildings, containing a multi-functional fitness center with big arena and various academic buildings, consume the majority of the campus energy and can represent common building type on a university campus. The amount of personnels inside different buildings is influenced by various factors. They are the maximum building capacity, the type of building and the event that is held inside the building. According to the events and schedules, the number of people in the five buildings are shown in Fig.~\ref{fig:amoutpeople}.

Bus $2$ is holds large events such as graduation commencements, any athletic games, the amount of people may vary a lot and the rate of change of population can fluctuate dramatically. Bus $59$ is the law school. It is a large building and holding relatively small conferences. Bus $41$ is a $4$ stories academic building. Bus $24$ is a buildings housing education and office activities. Bus $13$ contains many chemistry labs. We can demonstrate that our proposed methodology can be applied to different types of building and can handle dynamic optimization problems when the load and number of people are constantly changing. In this section, a typical summer day is selected, outdoor temperature is shown in Fig.~\ref{fig:amoutpeople}.

\begin{figure}[h]
	\centering
	\includegraphics*[width=0.7\linewidth]{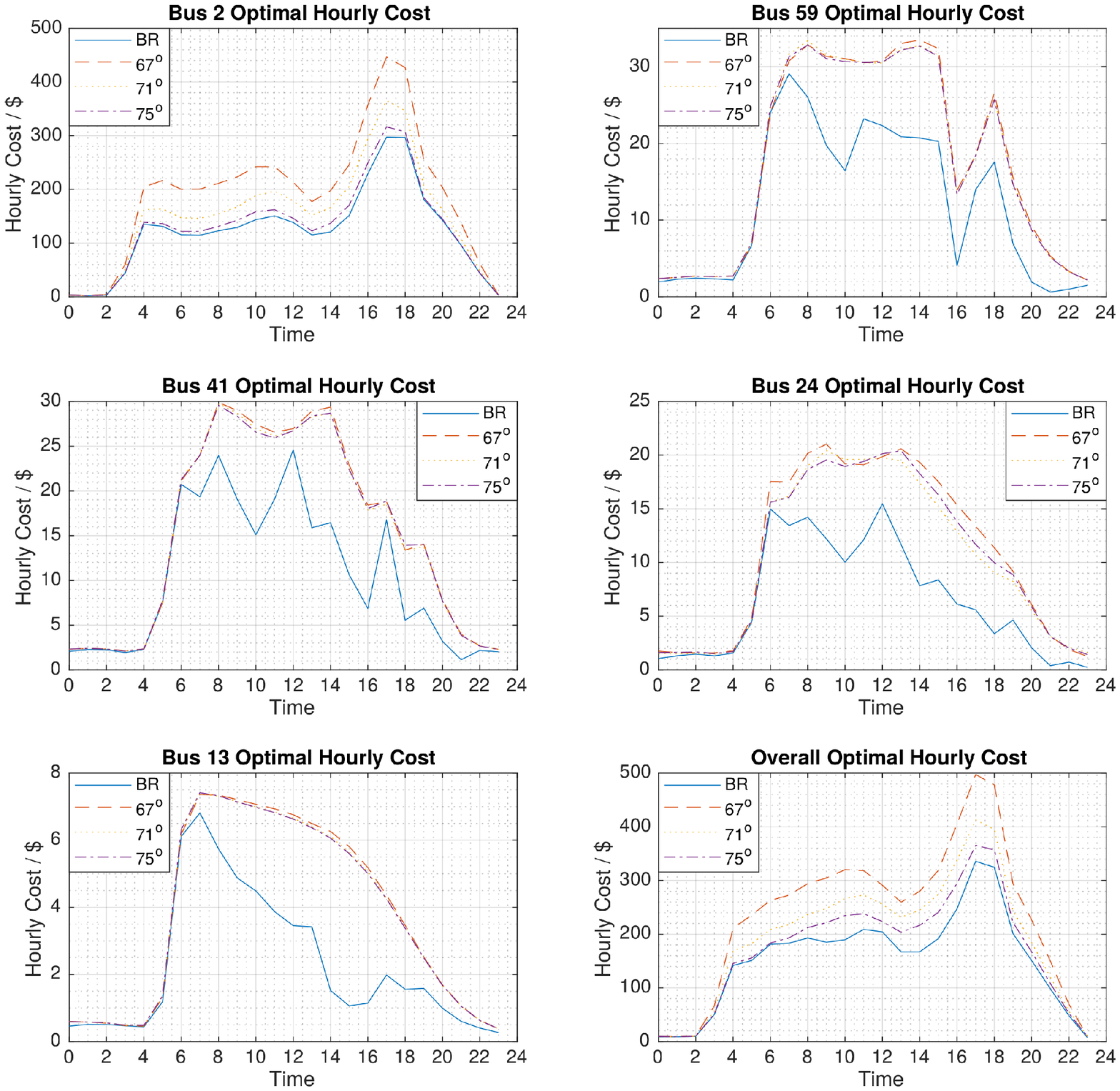}
	\caption{The results comparison}
	\label{fig:result}
\end{figure}

Fig.\ref{fig:result} shows the simulation results for each of the five selected buses and the overall social cost, respectively. For bus $2$, which is the largest buildings on campus, the daily indoor temperature is kept at $71\degree F$. The sole control strategy incurs $\$3,622.94$ in social cost. The best response strategy incurs $\$2,911.39$, which means that the proposed methodology can help the manager to save up to $\$711.5$ on the test day. Bus $13$ has the least square feet and capacity of all, the constant temperature control strategy incurs $\$89.6$ if the HVAC system is set at $75\degree F$. Compared to the optimal strategy that incurs $\$53.498$, the constant controlling method would make the manager pay $67\%$ more. For all the five selected buses, the total optimal daily cost is $\$3,655.6$ at the Nash equilibrium point, while the cost for $67\degree F$, $71\degree F$, and $75\degree F$ are $\$5,686.0$, $\$4,744.0$, and $\$4,215.2$, respectively. The simulation results prove that the Nash equilibrium point calculated by the proposed methodology is the global optimal point for the problem of social cost minimization. The outcome of this proposed methodology substantiates the use of it in buildings with central HVAC systems, by adaptively calculating the optimal HVAC system control settings based on DLMP and occupancy.

\section{Multi-Agent Reinforcement Learning for Social Cost Game}\label{sec:RL}
The social cost game depicted in Sec.~\ref{sec:game} is limited by the algorithm complexity. When the number of indoor temperature control strategy or the number of distribution node increase, the computation time increases exponentially. The shortcoming makes the cost of social energy game consume long time, although the modern computational capability of CPU or GPU is much better than the past. Hence, there is a need to implement an advanced algorithm to simplify the game strategy pool and adapt to the constantly changing environment. Even if the game control strategy or the node distribution power system goes up, the computational time should increase linearly not exponentially. Therefore, we implement the Markov Decision Process based Multi-Agent Reinforcement Learning to cope with the puzzle.
\subsection{Markov Decision Process for Social Energy Cost Game}\label{sec:MDP}
The framework of the MDP has the following components: (1) state of the system, (2)actions in each state, (3) transition probabilities, (4) transition rewards for each action from former state to next state, and (5) a policy. As demonstrated in the previous chapters and sections, the intelligent artificial model we build can model the system. So we can observe the Markov chain by observing the system model. The ideas are explained below.

\begin{itemize}
	\item State: For each player in the distribution power system who plays the social cost game, the "state" of a system is usually a set of measurable factors that can be used to describe their cost. In our case, the system is described in terms of the cost of social energy that consists the energy cost and the cost of working efficiency reduction, calculated by DLMP multiplying energy usage and indoor occupants multiplying the reduction of working efficiency, respectively. The system is a dynamic system that means any subtle change can incur the change of state.
	\item Action: The actions for MDP is the strategies for each player. The strategies allowed in the social cost game are the actions in the Markov Chain. The change of state can incur the change of action, vice versa.
	\item Transition Probability: Assume that in state $i$ action $a$ is selected to be the optimal game strategy, and state $j$ is the next state. Let $p(i,a,j)$ denotes the probability of transferring from state $i$ to state $j$ under the influence of action $a$.
	\item Immediate Rewards: The player receives an instant reward (which could be either positive or negative) when it transitions from one state to another, which is represented by $r(i,a,j)$.
	\item Policy: The policy defines the actions (strategy) to be chosen in every state visited by the system.
	\item Time of Transition: Some researches also introduce time of transition to their study. In our case, it is assumed that the time of transition is unity, which means that the transition time is not a factor during the study of MDP in this manuscript.
\end{itemize}

In our study, discounted reward is introduced to define the reward that the corresponding action (strategy) can make for the player. $x_s$ denote the state of the system before the $s$th transition. The objective of the discounted-reward MDP is to find the strategy that optimize the discounted reward starting from every state. The discounted reward from state $i$ can be defined as:

\begin{equation}
	\zeta_i = \lim_{k \rightarrow \infty} E [\sum_{s=1}^{k} \tau^{s-1} r(x_s,\pi(x_s),x_{s+1}) \mid x_1 = i]
\end{equation}\label{equ:MDP1}

where $\gamma$ denotes the discount factor, and $0 \leq \tau \leq 1$, an alternative expression of Equ.~\ref{equ:MDP1} is:

\begin{equation}
	\zeta_i = E[r(x_1,\pi(x_1),x_2) + \tau r(x_2,\pi(x_2),x_3) + \tau^2 r(x_3,\pi(x_3),x_4) + \cdots]
\end{equation}\label{equ:MDP2}

$\tau$ is used to discount the monetary value of the cost of social energy, and it should be noted that:

\begin{equation}
	\tau = (\frac{1}{1+\mu})^1
\end{equation}\label{equ:dcount}

where $\mu$ is the rate of interest. When $\mu>0$, we can ensure that $0 \leq \tau \leq 1$. In our study, it is assumed that the discounting rate is fixed so the power of $\frac{1}{1+\mu}$ is kept at $1$.

\subsection{The Implementation of Multi-Agent Reinforcement Learning for Social Cost Game}
With the definition of Markov Decision Process (MDP) in Sec.~\ref{sec:MDP}, we can implement a multi-agent RL algorithm that can be used to reduce the computational complexity for the proposed game in Sec.~\ref{sec:game}. It should be noted that the learning process of RL algorithm needs the updating of rewards in its database every time the system transition into a new state. Like in other researches that relates to RL algorithm, we define the constantly updating quantities as $Q$ factors as well. So $Q(i,a)$ will denote the reward quantity for state $i$ and action $a$.

The reward that is calculated in the transition is denoted as feedback. The feedback is used is to update the $Q$-factors for the evaluation of actions (strategies) in the former state. Generally speaking if the value of a feedback is good, the $Q$-factor of that action is increased, otherwise, the the $Q$-factor of that action is decreased. Therefore, the system is analyzed and controlled in real time. In each state visited, some action is opted out and the system is ready to proceed to next state. It should be noted that the "state" in our context is the power system condition at the specific time that the state is visited. Since at a specific time point, the number of indoor occupant is fixed, the only factor that influences the choice of HVAC setting is the DLMP at each bus. Each time the action is selected or changed, the DLMP will be influenced. Then, the system enters a new state.

\subsubsection{Discounted Reward Multi-Agent Reinforcement Learning for Social Cost Game}
In our study, we choose the discounted reward Multi-agent RL for our cost of social energy game to realize our goal of computational complexity reduction \cite{Rlearning}. The general steps for discounted reward Multi-agent RL can be expressed as follows:

\begin{itemize}
	\item Step 1 (Input and Initiation): Set the $Q$-factors to 0:
	\begin{equation}
		Q(i,a) \leftarrow 0, \forall i, and, \forall a
	\end{equation}
	$A(i)$ denotes the set of actions in state $i$. In our case, the number of action equals to the number of strategy in the social energy game. $\alpha^k$ defines the learning rate in the $k_{th}$ iteration. Set $k = 1$ when transition to a new state. $Itmax$ denotes the maximum number of iteration, and should be set to a large number. 
	\item Step 2 ($Q$-factor Update): Let $\mid A(i) \mid$ denotes the number of actions in set $A(i)$. Hence, the probability of action $a$ is selected in state $i$ is $\frac{1}{\mid A(i) \mid}$. $r(i,a,j)$ denotes the transition reward. The algorithm for updating $Q(i,a)$ is defined as:
	\begin{equation}
		Q(i,a) \leftarrow (1 - \alpha^k)Q(i,a) + \alpha^k[r(i,a,j) + \tau \max\limits_{b \in A(j)}Q(j,b)],
	\end{equation}
	The computation of $\alpha^k$ will be discussed later. $\tau$ denotes the discount factor in MDP.
	\item Step 3 (Termination Check): Increase $k$ by $1$. Set $i \leftarrow j$, when $k < Itmax$, then return to Step 1. Otherwise, proceed to Step 4.
	\item Step 4 (Outputs): For each state $i$, select the action $a^*$ that the corresponding $Q(i,a^*)$ achieves the optimal value. 
\end{itemize}

The learning rate $\alpha^k$ should be positive value and satisfy $\alpha^k<1$. The learning rate for the discounted reward reinforcement learning is a function of $k$ and have to meet the condition in \cite{RLcondition}. In our research, the learning rate step size is expressed as:
\begin{equation}
	\alpha^k = \frac{C}{D+k}
\end{equation}
where $C = 90$ and $D=100$ in our tests.

According to the general steps, we can formulate our discounted reward multi-agent RL social energy game as follows:
\begin{itemize}
	\item Step 1 (Input and Initiation): Set the $Q$-factors to 0:
	\begin{equation}
	Q(i,s) \leftarrow 0, \forall i, and, \forall s
	\end{equation}
	$S(i)$ denotes the set of strategy in game $\Psi$. In our case, the number of action equals to the number of strategy in the social energy game. $\alpha^k$ defines the learning rate in the $k_{th}$ iteration. Set $k = 1$ when transition to a new state. $Itmax$ denotes the maximum number of iteration, and should be set to a large number. In our research, the $Itmax = 10000$.
	\item Step 2 ($Q$-factor Update): Let $\mid S(i) \mid$ denotes the number of actions in set $S(i)$. Hence, the probability of strategy $s$ is selected in state $i$ is $\frac{1}{\mid S(i) \mid}$. $\delta(i,a,j)$ denotes the transition reward of the corresponding strategy. The algorithm for updating $Q(i,a)$ is defined as:
	\begin{equation}\label{eq:rl1}
	Q(i,a) \leftarrow (1 - \alpha^k)Q(i,a) + \alpha^k[\delta(i,s,j) + \tau \max\limits_{b \in S(j)}Q(j,b)],
	\end{equation}
	It should be noted that  $\max\limits_{b \in S(j)}Q(j,b)$ equals the optimal social cost $\gamma^{i,t}$ in game $\Psi$. Therefore, we can transform Equ.~\ref{eq:rl1} into:
	\begin{equation}
		Q(i,a) \leftarrow (1 - \alpha^k)Q(i,a) + \alpha^k[\delta(i,s,j) + \tau \gamma(i)],
	\end{equation}
	where $\gamma(i)$ denotes the optimal social energy game payoff.
	\item Step 3 (Termination Check): Increase $k$ by $1$. Set $i \leftarrow j$, when $k < Itmax$, then return to Step 1. Otherwise, proceed to Step 4.
	\item Step 4 (Outputs): For each state $i$, select the strategy $s^*$ that the corresponding $Q(i,a^*)$ achieves the optimal value.
	\item Pop up the best two strategy according to the optimal $Q(i,a^*)$ to play the social energy game
\end{itemize}

\subsection{Results}
In this chapter, bus $2$, bus $59$, bus $41$, bus $24$, and bus $13$ are selected as the social cost game players and are highlighted in Fig.\ref{fig:topology}. Those are the same five buildings as in Sec.~\ref{sec:gamer}. Since the building information has already been introduced in Sec.~\ref{sec:gamer}, it would not need to spend space do it twice. The number of people in the five buildings are shown in Fig.~\ref{fig:amoutpeople}.
\begin{figure}[!h]
	\centering
	\includegraphics*[width=1.0\textwidth]{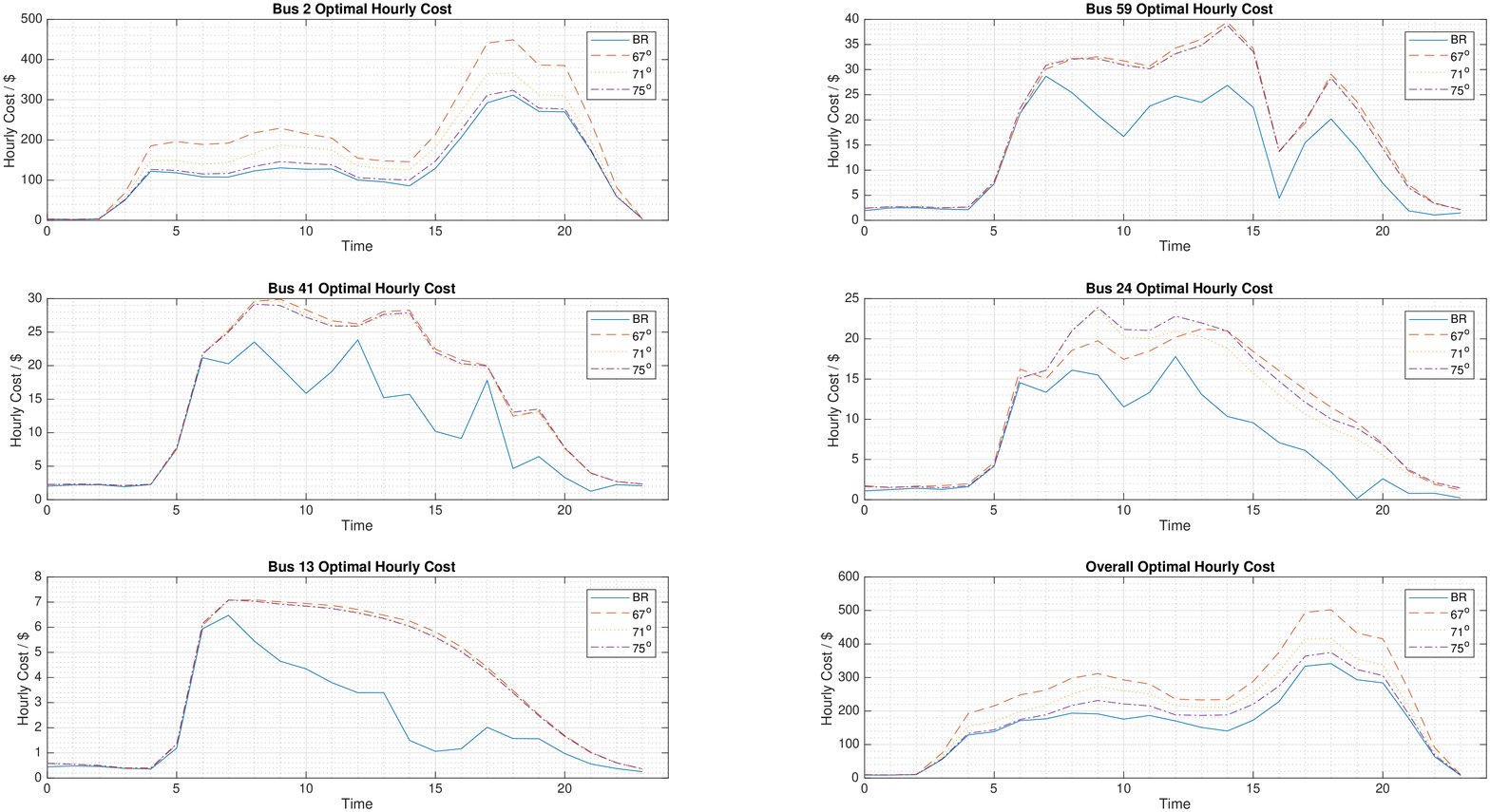}
	\caption{The summer day results comparison}
	\label{fig:resultRL1}
\end{figure}

\begin{figure}[!h]
\centering
\includegraphics*[width=1.0\textwidth]{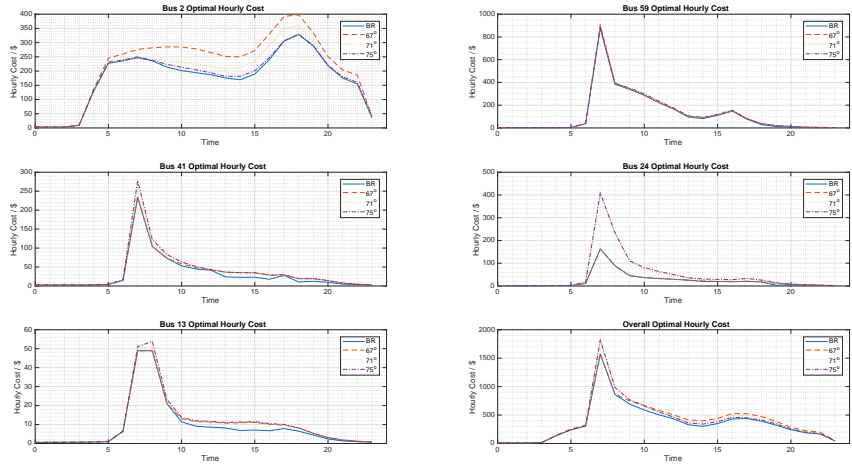}
\caption{The winter day results comparison}
\label{fig:resultRL2}
\end{figure}
The result in Fig.~\ref{fig:resultRL1} and Fig.~\ref{fig:resultRL2} shows that the proposed algorithm can achieve the same optimal outcomes as the algorithm we demonstrated in Sec.~\ref{sec:game}.

\begin{figure}[!h]
	\centering
	\includegraphics*[width=1.0\textwidth]{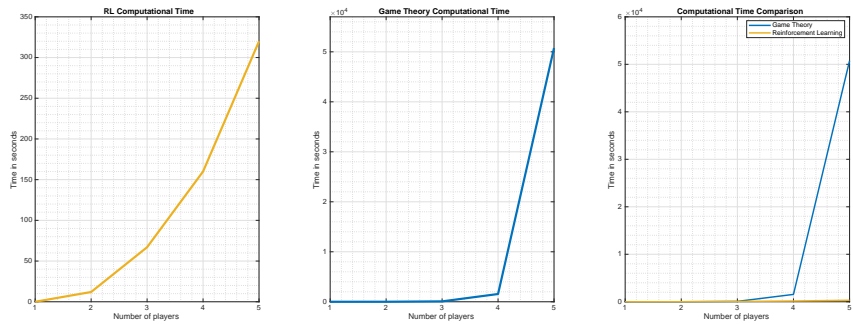}
	\caption{The computational time comparison}
	\label{fig:time}
\end{figure}

Fig.~\ref{fig:time} shows the computational complexity comparison between the two proposed methodologies. It shows that when the number of players in a game increases, the time complexity of the reinforcement learning based game can be approximated as a linear function, while the time complexity of the game theory can be approximated as an exponential function. When there are just two or three players in a game, the computation time for the proposed methodology in Sec.~\ref{sec:game} is smaller than the methodology depicted in Sec.~\ref{sec:RL}. As the number of players increases, the reinforcement learning based algorithm shows its undefeatable advantage compared with the algorithm based on game theory solely. 

\section{Chapter Conclusion}\label{sec:conclusion}
In this chapter, we proposed an autonomous and optimal HVAC system energy consumption scheduling algorithm for minimizing the social cost. That is the parameter we proposed to quantify and measure both human-bing working productivity and energy bill as the monetary value. Firstly, we implemented a realistic real-time pricing mechanism in our parallel system instead of utilizing simple pricing assumptions such as flat rate pricing. Unlike most of the existing game strategies that concentrate on the reduction of energy consumption and monetary cost, our methodology seeks to balance the energy cost and indoor air quality perceived by people. Our proposed algorithm is based on the interactions among players and the interaction between the energy-end users and power suppliers, and each of the players aims to maximize its own profit in a game-theoretic way. Simulation results prove that the proposed HVAC management strategy can reduce energy cost and also maintain the indoor working efficiency in a comfort zone. Second, to address the drawback of the proposed game-theoretic methodology, we implement reinforcement learning to reduce the computational complexity. This methodology can achieve the same optimal results but in much shorter time window.

%

\section{Proof of theorem 1}
In light of Lemma 1, the feasible set of $(\Psi)$ is nonempty as for every finite non co-operative game a Nash equilibrium exists. Further let $S$ be the feasible region for $(\Psi)$ then,
\begin{equation}
\text{min}_{(\alpha,\gamma^1,\dots,\gamma^n)\in S}\sum_{i \in N}(\gamma^i-\sum_{j=1}^{m^i}\delta^i(\alpha^{-i},s^i_j)) \geqslant 0.
\end{equation}
Thus, if $\alpha^*$ is a Nash equilibrium it is feasible for $(\Psi)$, and from (1),
\begin{equation}
\sum_{i \in N}(\gamma^{i*}-\delta^i(\alpha^*))=0
\end{equation}
yielding that $\alpha^*$ is an optimal solution of $(\Psi)$.

Conversely, suppose $(\alpha,\gamma^{1*},\dots,\gamma^{n*})$ is an optimal solution of $(\Psi)$ then it satisfies (2) to (4).

By virtue of (2), $\sum_{i \in N}(\delta^i(\alpha^{-i*},s^i_j))$.

But by the existence theorem of Nash equilibrium, there must exist at least on $(\alpha,\gamma^1,\dots,\gamma^n)$ feasible for $(\Psi)$ with $\sum_{i \in N}(\gamma^{i}-\delta^i(\alpha))=0$. So for $(\alpha,\gamma^{1*},\dots,\gamma^{n*})$ to be a blobal minimum for $(\Psi)$,
\begin{equation}
\sum_{i \in N}(\delta^i(\alpha^*)-\gamma^{i*})=0
\end{equation}

Consequently $\gamma^*$ is a Nash equilibrium of game $\psi$, on account of Lemma 1 the payoff $\delta^{i*}$ is obviously the optimal expected payoff to player $i$.

We see that the problem of computing a Nash equilibrium of $\psi$ reduces to that of solving the optimization problem $(\psi)$ with optimal value zero.

\chapter{Chapter 6}\label{Chap6}
\section{Conclusion}
Since the invention of power system, power engineers always aim to serve human-beings better quality and more stable energy. In this manuscript, the research investigates the mutual influences between the building energy usage and the productivity of indoor occupants. The preliminary results reveal that there is a strong relationship between those two factors. Our research aims to find the optimal balance between building energy usage and the working efficiency of indoor occupants. To address the novel distribution power system problem, we implement a parallel computing scheme and define the "social energy" as a novel concept that combines the monetary and society aspects of power system together. The parallel computing scheme helps us to build the interactive mechanism between physical system and artificial system. Then, the study and research of the aforementioned problem is conducted from shallow to deep. The investigation starts with the implementation of techniques such as quadratic regression model and brute search algorithm. Distribution locational marginal pricing is also introduced to calculate the economic cost of energy. Then, to cope with the future smart grid features, game theory is implemented. But limitation of the game theory is obvious. With the increase of bus or strategy, the computation time goes up quickly. Therefore, reinforcement learning is implemented and embedded into the game-theoretic methodology to reduce the computational complexity. The results reveal that the algorithm achieves the time reduction goal and simplify the gaming process and can still optimize the indoor temperature control strategy. Our research can help to achieve a better demand side management with the consideration of indoor occupants and a better power delivery quality.
\newpage
\addcontentsline{toc}{chapter}{References}
\renewcommand{\bibname}{References}
\bibliographystyle{IEEEtran} 
\bibliography{ComprehensiveH}


\end{document}